\documentclass{article}
\usepackage{latexsym}

\begin{document}
\begin{center}

\Large\bf {Spin-dependent tunneling and Coulomb blockade in ferromagnetic nanoparticles}
\end{center}
\begin{center}
\bf{Kay Yakushiji}$^{1\ast }$\textbf{, Seiji Mitani}$^{1}$\textbf{, Franck Ernult}$^{1}$\textbf{, \newline
Koki Takanashi}$^{1}$\textbf{, Hiroyasu Fujimori}$^{2}$
\end{center}

\begin{center}
$^{1}$Institute for Materials Research, Tohoku University, \newline
$^{2}$The Research Institute for Electronic and Magnetic Materials \newline

\end{center}

$^{*}$Corresponding author: 

Spintronics Group, AIST

Tsukuba Central 2, Umezono 1-1-1, 
Tsukuba 305-8568, Japan.

k-yakushiji@aist.go.jp\newline
\newline

\textbf{Abstract}\newline
In this paper we review studies on spin-dependent transport in systems 
containing ferromagnetic nanoparticles. In a tunnel junction with a 
nanometer-scale-island, the charging effect leads to an electric current 
blockade phenomenon in which a single electron charge plays a significant 
role in electron transport, resulting in single-electron tunneling (SET) 
properties such as Coulomb blockade and Coulomb staircase. In a tunnel 
junction with a ferromagnetic nano-island and electrode, it was expected 
that the interplay of spin-dependent tunneling (SDT) and single-electron 
tunneling (SET), i.e., spin-dependent single-electron tunneling (SD-SET), 
would give rise to remarkable tunnel magnetoresistance (TMR) phenomena. We 
investigated magnetotransport properties in both sequential tunneling and 
cotunneling regimes of SET and found the enhancement and oscillation of TMR. 
The self-assembled ferromagnetic nanoparticles we have employed in this 
study consisted of a Co-Al-O granular film with cobalt nanoparticles 
embedded in an Al-O insulating matrix. A Co$_{36}$Al$_{22}$O$_{42}$ film 
prepared by a reactive sputtering method produced a TMR ratio reaching 10 
{\%} and superparamagnetic behavior at room temperature. The TMR ratio 
exhibited an anomalous increase at low temperatures but no indication of 
change with bias voltage. In Ch. 4, we show that the anomalous increase of 
the MR provided evidence for higher-order tunneling (cotunneling) between 
large granules through intervening small granules. We emphasize that the 
existence of higher-order tunneling is a natural consequence of the granular 
structure, since broad distribution of granule size is an intrinsic property 
of granular systems. In Ch. 5, we concentrate on SD-SET properties in 
sequential tunneling regimes. We fabricated two types of device structures 
with Co-Al-O film using focused ion-beam milling or electron-beam 
lithography techniques. One had a granular nanobridge structure: 
point-shaped electrodes separated by a very narrow lateral gap filled with 
the Co-Al-O granular film. The other had a current-perpendicular-to-plane 
(CPP) geometry structure: a thin Co-Al-O granular film sandwiched by 
ferromagnetic electrodes with the current flowing in the direction 
perpendicular to the film plane through a few Co particles. We found the 
enhancement and oscillation of TMR due to spin-dependent SET in sequential 
tunneling regimes. In Ch. 6, we report experimental evidence of a spin 
accumulation effect in Co nanoparticles leading to the oscillation of TMR 
with alternate sign changes. Furthermore, we discovered that the spin 
relaxation time in the nanoparticles is unprecedentedly enhanced up to the 
order of more than hundreds of nanoseconds, compared to that evaluated from 
the spin-diffusion length of ferromagnetic layers in previous CPP-GMR 
studies, $i.e., $the order of tens of picoseconds.

\newpage
\textbf{Table of contents:}\newline

\textbf{1. Introduction }\newline

\textbf{2. Basic concepts}

2.1 Charging effect of nanoparticles

2.2 Spin-dependent single-electron tunneling

2.3 Spin accumulation in nanoparticles\newline

\textbf{3. Insulating granular films }

3.1 Materials, sample preparation and microstructure

3.2 Electric transport and magnetic properties in insulating granular 
systems

3.3 Particle size distribution in insulating granular systems\newline

\textbf{4. Spin-dependent higher order tunneling in Coulomb blockade regime}

4.1 Results

4.2 Theoretical explanation

4.2.1 Temperature dependence of MR

4.2.2 Bias-voltage dependence of MR \newline

\textbf{5. Spin-dependent single-electron tunneling in microfabricated 
structures}

5.1 Current path restriction to observe single-electron tunneling in 
sequential tunneling regime

5.2 Results in nanobridge structures

5.2.1 Experimental procedures and results

5.2.2 Theoretical explanation

5.3 Results in CPP geometry structures\newline

\textbf{6. Spin accumulation effect in nanoparticles}

6.1 Experimental procedures and results

6.2 Numerical analysis

6.3 Evaluation of spin relaxation time\newline

\textbf{7. Summary}

\newpage 
\textbf{1. Introduction}\newline

Ever Since the giant magnetoresistance (GMR) effect was discovered in 
1988[1], magnetotransport phenomena in magnetic multilayers 
with precise layer-by-layer deposition on a nanometer scale have attracted 
much attention due to the potential applications of large magnetoresistance. 
GMR and Tunnel magnetoresistance (TMR)[2-4] discoveries have 
raised physical issues regarding the interplay between the spin and charge 
of an electron. From the industrial point of view, the control of this 
interplay is important because larger magnetoresistance is key to higher 
performance of data storage applications[5-8]. 

Progress in microfabrication technique has enabled us to study 
characteristic phenomena in laterally small sizes, and it has long been 
known in metallic or semiconductor systems, nanometer-sized structures 
provide insight into mesoscopic phenomena such as quantization of 
conductance, single-electron tunneling, discreteness of the chemical 
potentials, and so on. To study a mesoscopic property, we must reduce the 
size of a structure within a characteristic length according to its feature 
and temperature. For example, to study Coulomb blockade and resulting 
single-electron tunneling (SET) properties at room temperature, the size of 
a particle should be less than 3 nm (embedded in a vacuum) [9]. A 
microstructure a few nanometer in size, which is under the microfabrication 
limit, is generally prepared by a bottom-up process. 

Returning to the magnetic systems, and in view of GMR or TMR in layered 
nanostructures, it has naturally been anticipated that spin-dependent 
transport in three-dimensional magnetic nanostructures would gives rise to 
further novel phenomena. Our interest emerged especially in the interplay of 
spin-dependent tunneling (SDT) and SET in magnetic nanoparticles. TMR, which 
was discovered in 1995[3, 4], has normally been studied in 
magnetic tunnel junctions (MTJs) consisting of a 1-3 nm-thick insulating 
barrier layer sandwiched between upper and lower ferromagnetic electrodes, 
and its ratio is given by the difference in the electrical resistance 
between parallel and anti-parallel alignments of magnetization vectors. The 
TMR ratio of a macroscopic MTJ with an amorphous insulating barrier is 
almost consistent with that proposed by Julliere [10] in terms of the spin 
polarizations of the electrodes. In the case that the size of the MTJs is 
reduced to nanometer-scale, the interplay of SDT and SET causes a 
peculiarity in the TMR ratio different from that of the Julliere model. 

Here we report characteristic magnetotranport behaviors arising from the 
interplay of SDT and SET in samples using magnetic nanoparticles. We 
employed an insulating granular film consisting of a few nanometer-sized 
magnetic particles embedded in an insulating matrix. The size of the 
particles was so small that their charging energy exceeds 30 meV ($\sim $300 
K), in which case SET is expected to occur between pairs of neighboring 
particles at low temperatures. In a granular film of macroscopic size 
containing a large number of particles, however, SET phenomena between 
neighboring particles average out, due to the large distribution of particle 
sizes and interparticle distances. In order to observe SET phenomena such as 
Coulomb staircase and Coulomb blockade with a clear threshold, the tunnel 
paths must be restricted from among the vast number of possible paths by 
uniting them with microfabricated electrodes.

First, we describe characteristic magnetotransport behavior in 
non-microfabricated granular films. Although SET properties average out in 
non-microfabricated samples, the higher-order tunneling process in a Coulomb 
blockade regime, or so-called cotunneling, gives rise to enhanced TMR at a 
low-temperature and low-bias-voltage region. Next, we describe 
magnetotranport behavior for microfabricated granular films. The 
microstructures using granular film we prepared in this study consisted of 
either a granular nanobridge or a granular CPP 
(current-perpendicular-to-plane) structure. In the former, or lateral-type 
microstructure, a granular film was placed in a nanogap of microfabricated 
electrodes. We found enhanced TMR associated with Coulomb blockade in this 
structure. The latter was a nanopillar of layered structure: in this case, a 
granular film was sandwiched between electrodes. In the CPP structure, we 
found interesting TMR behavior due to SET and spin accumulation in 
nanoparticles; in addition, it was elucidated that spin relaxation time is 
much enhanced in the Co nanoparticles. 

\newpage 
\textbf{2. Basic concepts}\newline

2.1 Spin-dependent tunnel magnetoresistance\newline

Tunnel magnetoresistance (TMR) originates from spin-dependent tunneling in 
an FM / I / FM magnetic tunnel junction (MTJ), where the FMs are 
ferromagnetic layers and the I is a tunnel barrier that is a few nm thick. 
In 1975, Julli\'{e}re formulated a model for the difference in conductance 
between the parallel and anti-parallel configurations in the two FM 
electrodes, FM1 and FM2. TMR is defined as:
$$
{\mbox{TMR}}=\frac{\Delta R}{R_P }=\frac{R_{AP} -R_P }{R_P }=\frac{G_P -G_{AP} 
}{G_{AP} },
\eqno(2.1) $$
where $R_{P(AP)}$ is the resistance and $G_{P(AP)}$ is the conductance for the 
P (AP) alignment [10]. $G$ is given from the sum of the $G$ for each spin channel 
$\sigma $, and $G_{\sigma }$ is described by
$$
G_\sigma \propto \sum {D_{1\sigma } D_{2\sigma } } ,
\eqno(2.2) $$
where $D_{1(2)\sigma }{\rm g}$s the density of states of the $\sigma $ spin 
channel at the Fermi level in FM1(FM2). The total conductance in the P 
alignment is given by
$$
G_P =G_P^\uparrow +G_P^\downarrow \propto D_{M1} D_{M2} +D_{m1} D_{m2} ,
\eqno(2.3) $$
and in the AP alignment
$$
G_{AP} =G_{AP}^\uparrow +G_{AP}^\downarrow \propto D_{M1} D_{m2} +D_{m1} 
D_{M2} ,
\eqno(2.4) $$
where $D_{Mi}$ and $D_{mi}$ are the densities of states for the majority and 
minority spin bands in the $i$-th FM electrode, respectively. The spin 
polarization of the $i$-th electrode is defined by
$$
P_i =\frac{D_{Mi} -D_{mi} }{D_{Mi} +D_{mi} }.
\eqno(2.5) $$
Using this definition, Eq. (2.1) can be expressed as
$$
{\mbox{TMR}}=\frac{G_P -G_{AP} }{G_{AP} }=\frac{2P_1 P_2 }{1-P_1 P_2 }.
\eqno(2.6) $$
According to Eq. (2.6), which was predicted by Julli\'{e}re, a TMR of about 
28 {\%} is expected assuming both electrodes are Co with P $\sim $ 0.35. 
This prediction of the Julli\'{e}re model almost exactly reflects the 
experimental results for MTJs with Al-O amorphous barriers. 

In the case of insulating granular systems, we normally define the TMR ratio 
as TMR = $\Delta R/ R_{max}$. The TMR ratio should be smaller than those of 
MTJs because of the random orientation of the magnetization ($M)$ at the zero 
field resulting in an incomplete anti-parallel configuration. Inoue and 
Maekawa considered the relative magnetization of a granular system $m=M$ / 
$M_{s}$, where $M_{s}$ is the saturation magnetization, and derived the TMR 
as [11]:
$$
{\mbox{TMR}}=\frac{m^2P^2}{1+m^2P^2}.
\eqno(2.7) $$
Assuming Co nanoparticles with P$\sim $0.35, the saturated ($m$ = 1) TMR ratio 
$P^{2}$/(1+$P^{2})$ is evaluated to be about 11 {\%}.\newline

\textbf{2.1 Charging effect of nanoparticles}\newline

The charging effect of nanoparticles leads to a Coulomb blockade; then the 
single-electron charge plays a significant role in electron transport [9, 
12, 13]. A simplified model is shown in Fig. 2-1: an individual particle, 
isolated in an insulating barrier, is located between source and drain 
electrodes, and the tunneling electron travels from the source to the drain 
through the particle. The tunneling of electrons can be inhibited, and the 
current does not flow at small bias voltages if the electrostatic energy 
$e^{2}$/2$C$ of a single excess electron on the island is much larger than the 
thermal energy $k_{B}T$, where $C$ is the capacitance of the island. The 
suppression of current at small bias voltages is called a Coulomb blockade. 
When the bias voltage increases and exceeds the threshold $V_{th} =e/2C$, 
the current starts to increase. If one junction resistance is similar to 
that of other ($R_1 \approx R_2 )$, the current increases smoothly with bias 
voltage (see Fig. 2-2 (a)). On the other hand, if the difference between the 
two junction resistances is very large ($R_1 <<R_2 $or $R_1 >>R_2 )$, the 
current increases stepwise with bias voltage depending on the number of 
electrons accumulated on the island (see Fig. 2-2 (b)). The step-like 
structure in current vs. bias-voltage ($I-V_{b})$ characteristics is called 
the Coulomb staircase. The Coulomb blockade and the Coulomb staircase are 
representative phenomena of single-electron tunneling (SET). In practice, 
even for multiple junctions including not only one but some islands between 
electrodes, SET phenomena can be observed. 

For the appearance of SET phenomena, the following two conditions are 
generally essential. First, as mentioned above, the charging energy 
$E_{c}$ required to add an electron to an island must far exceed the thermal 
energy; i. e.,
$$
E_c >>k_B T.
\eqno(2.8) $$
Second, the junction resistance $R_{J}$ must exceed the resistance quantum 
$R_{K}=h$/$e^{2}\approx $25.8 k$\Omega $; i. e., 
$$
R_J >>R_K ,
\eqno(2.9) $$
which ensures that the wave function of an excess electron on an island is 
localized there. Without this condition, the electron would be delocalized 
through the island, permitting the transport. SET phenomena in nonmagnetic 
systems, where both islands and electrodes are nonmagnetic, have been 
extensively investigated, both theoretically and experimentally [9]. The 
islands must be small enough to satisfy Eq. (2.8), since $E_{c}$ increases as 
the island size decreases. The $E_{c}$ of an isolated island is described as
$$
E_c =e^2/4\pi \varepsilon d,
\eqno(2.10) $$
where $d$ is the diameter assuming the island is a sphere, and $\varepsilon $ 
is the dielectric constant of an insulator around the island. Although 
$E_{c}$ is somewhat modified by the configuration of surrounding islands and 
electrodes in a real system, we may consider that $E_{c}$ is roughly 
proportional to 1/$d$. The $E_{c}$ for submicron-sized islands is generally of 
the order of 10$^{-4}$-10$^{-3}$ eV (1$\sim $10 K). Thus, SET phenomena can 
be observed only at very low temperatures.

There are two distinct tunneling processes by which electrons are 
transferred between the electrodes via a small island: sequential tunneling 
and cotunneling. In sequential tunneling, there is no correlation between 
tunneling events into and out of the island. In this process, electron 
tunneling in either of the two junctions causes an increase of charging 
energy, and is suppressed by a Coulomb blockade. However, cotunneling occurs 
when tunnel resistance is not so high (1 or 2 orders higher) compared to the 
resistance quantum ($R_{K}\sim $ 25.8 k$\Omega )$. In the cotunneling 
process, two electrons tunnel in a correlated fashion, i.e., an electron 
tunnels into the island while the second electron simultaneously leaves the 
island through the other junction; the island is only virtually charged in 
this cotunneling process, and therefore there is no increase in charging 
energy in the overall tunneling process. Therefore, the cotunneling process 
contributes to a finite current even in a Coulomb blockade regime.\newline
\newline
\textbf{2.2 Spin-dependent single-electron tunneling}\newline

In the last section, we neglected the spin degree of freedom. In the 
magnetic tunnel junctions, the tunnel current depends on the relative 
orientation of magnetizations in the ferromagnetic electrodes. Assuming a 
double tunnel junction model with a ferromagnetic nanoparticle put between 
ferromagnetic electrodes, it was expected that the interplay of 
spin-dependent tunneling (SDT) and single-electron tunneling (SET), so 
called spin-dependent single-electron tunneling (SD-SET), would give rise to 
remarkable tunnel magnetoresistance (TMR) phenomena. 

Magnetotransport properties have been studied in both sequential tunneling 
and cotunneling regimes. In the limit of the sequential regime, phenomena 
are discussed in the framework of the orthodox theory of SET. Pioneering 
theoretical studies were performed by Barnas and Fert [14, 15], and 
Majumdar and Hershfiled [16] in 1998. They predicted that novel TMR 
behaviors such as modification of the spin-dependent tunneling probability 
associated with Coulomb staircase give rise to a TMR peak around the step 
point of the staircase. A peak appears at each step resulting in the 
oscillation of TMR as a function of the bias-voltage. Following their 
studies, further analyses were made by several authors [17-36]. Some 
studies have considered SD-SET with additional effects such as spin 
accumulation [17-30, 
36] and a discrete energy spectrum [20, 32-34] in the 
island. Characteristic TMR behavior due to the spin accumulation effect was 
predicted: the difference in the spin-splitting of the chemical potential 
between parallel and anti-parallel alignment of magnetic vectors gives rise 
to an alternate sign change of TMR [15, 27, 28, 30, 37]. The spin 
accumulation effect has been mainly considered in an SET device model with a 
nonmagnetic nano-island placed between ferromagnetic electrodes. Although 
TMR should not appear in the framework of the Julli\'{e}re model, 
nonequilibrium magnetic polarization due to spin accumulation leads to 
nonzero TMR with periodic oscillation. Studies concerning spin accumulation 
have also examined the magnetotransport properties for various spin 
relaxation times in the island because the accumulation occurs when the spin 
of an electron entering a particle does not flip until the next electron 
arrives.

On the other hand, in the case of the cotunneling process [38], 
magnetotransport phenomena have been analyzed in a Coulomb blockade 
region [36, 39-41]. In contrast to those in sequential tunneling, 
which occur outside of a Coulomb blockade regime and with in the limit of 
high tunnel resistance, cotunneling gives a dominant contribution when the 
tunnel resistances are no more than 1 or 2 orders higher than the resistance 
quantum. Assuming a double junction, the total tunnel resistance for 
sequential tunneling is proportional to the sum of the resistances, while 
that for cotunneling is proportional to their product. It has been predicted 
that this discrepancy would lead to enhanced TMR in a cotunneling regime 
 [41]. 

Experimentally, there have been several studies [42-72]. The first 
experimental evidence for characteristic SD-SET phenomena was observed by 
Ono et al. in 1995 [42]. They fabricated Ni / NiO / Co / NiO / Co 
double junctions with small contact area ($\sim $200nm$^{2})$ and measured 
magnetotransport properties by changing the gate voltage at a low 
temperature, below 1 K. In the Coulomb blockade region, they found enhanced 
TMR exceededing 40 {\%} in the off-state but only reacing 4 {\%} in the 
on-state. The enhancement was theoretically explained by a strong tunneling 
model in which the tunnel resistances were near the resistance 
quantum [39, 73]. Although the authors of these studies successfully 
observed enhanced TMR, the operating temperature was lower than 100 mK 
because the size of the microfabricated island was larger than 100 nm. 
Schelp et al also found a Coulomb blockade effect on TMR in a sample with 
nanometric Co clusters [55]. They prepared a layered Co / Al-O / 
Co clusters / Al-O / Co sample and observed twice-larger TMR at 4.2 K than 
that at RT. The origin for enhanced TMR was suggested to be the effect from 
the Coulomb blockade in Co clusters. Their observation of the Coulomb 
blockade and TMR even in the sample with a very large contact area was of 
significance, because they showed the potential of studying SD-SET in a 
layered structure. Following their study, several other SD-SET studies in 
layered structures have been 
performed [48, 53, 56, 58, 63, 65]. 
In order to understand the mechanism of SD-SET, as mentioned above, it is 
necessary to divide the properties of SD-SET into two categories according 
to transport mechanisms: cotunneling and sequential tunneling. Although 
there have been some reports on enhanced TMR ascribed to finite 
spin-dependent transport due to the cotunneling process, no experimental 
evidence in a sequential tunneling regime has been reported. We have 
fabricated appropriate sample structures using Co-Al-O insulating granular 
films, and elucidated SD-SET phenomena in sequential tunneling 
 [60, 63-65] and co-tunneling regimes [62, 
66] individually. Our study is the first to reveal an oscillatory TMR 
behavior in a sequential tunneling regime. The details are described later. \newline
\newline
\textbf{2.3 Spin accumulation in nanoparticles}\newline

Spin injection and accumulation were first studied in a layered sample where 
a normal metal (NM) layer was sandwiched between two ferromagnetic (FM) 
layers. A non-equilibrium effect of spin accumulation is generated by a 
spin-polarized current from one FM layer into the NM layer. When the NM 
layer is thinner than the spin-diffusion length, a spin injection signal 
attributed to spin accumulation in the NM layer is detected in the other FM 
layer according to the relative orientation of the two FM layers [74]. 
Spin accumulation has also been studied in sub-micrometer-sized lateral 
structures [75, 76]. In this manuscript, we discuss the 
spin accumulation effect in an isolated nanoparticle in which spin-polarized 
current is injected through a tunnel barrier. In such a system, spin 
accumulation in the nanoparticle occurs when the spin relaxation time 
overcomes the interval of successive electron tunneling 
 [15, 19, 21, 23, 25, 36, 37, 77, 78]. 
According to the sample structure shown in the Ch. 6, we introduce the FM / 
I / FM nanoparticle / I / NM model (I: insulating barrier) as illustrated in 
Fig. 2-3. As mentioned in Ch. 2, Sec.1, a tunnel conductance $G_{\sigma }$ 
for a spin states $\sigma {\rm g}$s expressed by the multiplication of the 
initial and final density of states (DOS) of the $\sigma $ spin band at 
Fermi level ($E_{F})$. There should be a discrepancy between incoming and 
outgoing numbers of spins at a fixed spin state because the DOS at $E_{F}$ of 
FM is spin-polarized while that of NM is not. This gives rise to the 
increase (or decrease) of spin population leading to the shift of the 
chemical potential $\Delta E_F^\sigma $ for a $\sigma $ spin state. In order 
to maintain charge neutrality in the particle, the chemical potential of 
another $\sigma $ shifts in the opposite direction, satisfying $D_\uparrow 
\Delta E_F^\uparrow =-D_\downarrow \Delta E_F^\downarrow $. Although the 
degree of the shift depends on the net DOS of the particle, $\Delta 
E_F^\uparrow $is not the same as $\Delta E_F^\downarrow $ because the net 
DOS is spin-polarized around $E_{F}$. In order to maintain spin-conservation 
conditions, the following expression is applied:
$$
\frac{(I_{1,\sigma } -I_{2,\sigma } )}{e}=\frac{D_\sigma \Omega }{\tau 
_{SF} }\Delta E_F^\sigma 
\eqno(2.11) $$
where $I_{i,\sigma } $ is the current at the $i^{th}$ junction ($i $= 1, 2) for 
spin $\sigma $, $\Omega $ is the volume of the particle, $\tau _{SF} 
$ is the mean spin relaxation time in a nanoparticle defined as $\tau 
_{SF}^{-1} =(\tau _\uparrow ^{-1} +\tau _\downarrow ^{-1} )/2$, and 
$\tau _\sigma $ is the spin relaxation time of electrons with spin $\sigma $. 
When we consider the spin accumulation effect in SET regime, the 
bias-voltage dependence of $\Delta E_F^\sigma $shows a sawlike oscillation 
with a period close to that of the Coulomb staircase. In the case of the Co 
nanoparticle, characteristic oscillation gives rise to modifications in the 
current -- bias voltage curve resulting in novel TMR behavior. Details will 
be shown in Ch. 6. 

\newpage 
\textbf{3. Insulating granular film}\newline

\textbf{3.1 Materials, sample preparation and microstructure }\newline

Insulating granular films consist of small metallic particles and an 
insulating matrix. If the composition of the metal is lower than the 
percolation limit, the transport is dominated by the tunneling of electrons 
between particles. If the particles are magnetic, TMR arises because the 
magnetization vectors on particles, which are not aligned at low applied 
fields, become more aligned as the applied field increases, leading to a 
decrease in resistivity. Pioneering works on TMR in insulating granular 
films were published by Gittleman et al. .[79] and by Helman and Abeles 
 [80], both of whom used Ni-Si-O films. However, the magnitudes of 
TMR were very small. In 1994 Fujimori et al. used a Co-Al-O granular film 
and reported a large TMR, which reached 10 {\%} even at rooom 
temperature [81]. This ratio was found to increase with decreasing 
temperature, exceeding 20 {\%} at low temperatures [62, 82]. 

Insulating granular films are easily prepared by sputtering and evaporation 
techniques. As shown in Fig. 3-1, the primary methods we used included (a) 
reactive sputtering, (b) sputtering with a composition target, and (c) 
tandem deposition with plural targets. We show the details of these methods 
for the preparation of a Co-Al-O granular film. In the case of (a), a Co-Al 
alloy target was sputtered in an Ar+O$_{2}$ atmosphere; then Al was 
selectively oxidized. In the case of (b), Al$_{2}$O$_{3}$ sheets are were 
placed over a Co target, and sputtering was performed in an Ar atmosphere. 
In the case of (c), there were two cathodes with targets, Co and 
Al$_{2}$O$_{3}$, and the rotation of the substrate holder caused the 
alternation of their deposition. In all cases, a granular structure 
comprising nanometer-sized Co metal particles embedded in an insulating Al-O 
matrix is formed on substrates. 

In this manuscript, we employed Co-Al-O films prepared by the reactive 
sputtering method. Hereafter the Co-Al-O granular film containing x at.{\%} 
Co, y at{\%} Al and z at.{\%} O is denoted Co$_{x}$Al$_{y}$O$_{z}$. The 
compositions are determined by Rutherford backscattering spectroscopy (RBS) 
analysis. The atomic composition of Co was roughly controlled by that in the 
Co-Al target. On the other hand, the composition of the film, at its fixed 
target composition, is relatively insensitive to the oxygen gas flow ratio. 
For example, in the case of the Co$_{36}$Al$_{22}$O$_{42}$ film, the target 
composition was fixed to be Co$_{25}$Al$_{75}$ while the oxygen gas flow 
ratio was allowed to range from 1.6 {\%} to 3.0 {\%} (at total gas pressure 
$\sim $ 1mTorr). 

Figs. 3-2 show (a) plan view and (b) cross-sectional transmission electron 
microscopy (TEM) micrographs for Co$_{46}$Al$_{19}$O$_{35}$ film [83]. 
The film has an isotropic granular structure consisting of Co particles 2-3 
nm in diameter (dark spheres) and intergranular Al-O about 1 nm in thickness 
(white channels). Fig. 3-3 shows a high resolution TEM micrograph for 
Co$_{52}$Al$_{20}$O$_{28}$ film, indicating that a crystalline Co particle 
is surrounded with non-stoichiometric aluminum oxide with amorphous 
structure. \newline

\textbf{3.2 Electric transport and magnetic properties in insulating 
granular systems}\newline

As mentioned above, the electric charging effect of small islands causes a 
Coulomb blockade when the charging energy overcomes the thermal and/or bias 
potentials. An insulating granular system consists of metallic nanoparticles 
embedded in an insulating matrix. Each nanoparticle is electrically 
isolated; their average size is a few nanometers while their charging energy 
exceeds 100 K. The electric transport in insulating granular systems 
represents a characteristic dependence on temperature. Fig.3-4 shows the 
temperature dependence of electrical resistivity\textit{${\rm g}\rho $} for Co-Al-O films of 
different compositions: Co$_{54}$Al$_{21}$O$_{25}$, 
Co$_{52}$Al$_{20}$O$_{28}$, Co$_{46}$Al$_{19}$O$_{35}$, and 
Co$_{36}$Al$_{22}$O$_{42}$. The plot of ln \textit{$\rho $} versus $T^{-1/2}$ is 
approximately linear for all the films. The relationship of $\ln \rho 
\propto 1/\sqrt T $was first derived by Sheng \textit{et al}. [84], considering 
electron tunneling, charging effect and particle size distribution. They 
analyzed its temperature dependence assuming a model in which particles on 
each contribution path have the same size $d$ and are separated by a barrier 
thickness $s$, keeping their ratio $s$/$d$ (or equivalently $E_{c}s)$ constant for a 
given composition. In this model, the electrical conductivity is dominated 
by tunneling between small particles at high temperature and between large 
ones at low temperatures because larger ones cause a Coulomb blockade as the 
temperature decrases.

For a conventional sandwich-type tunnel junction, the tunnel conductance is 
temperature independent and proportional to exp{\{}- 2(2$\pi $/$h)$(2\textit{m$\phi $})$^{1/2}s${\}}, 
where $h$ is the Planck's constant, $m$ the effective electron mass, \textit{$\phi $} the 
effective barrier height, and $s$ the barrier width. In the case of an 
insulating granular system, the existence of finite charging energy 
($E_{c})$ gives rise to the particular temperature dependence of $\rho $. The 
number density of the charge carrier, whose generation requires an 
$E_{c}$, is proportional to the Boltzmann factor 
exp{\{}$E_{c}$/2$k_{B}T${\}}, where $E_{c}$ is the Coulomb energy, $k_{B}$ the 
Boltzmann's constant, and $T$ the absolute temperature. The final formula about 
$\rho $ was obtained as

$$\rho \quad = \quad \rho _{0}{\{} \exp {\{}2( C/ k_{B})^{1/2}T^{-1/2}{\}},
\eqno(3.1)$$
where $C$ = (2$\pi $/$h)$(2\textit{m$\phi $})$^{1/2}$\textit{ sE}$_{c}$ which was named the tunnel activation 
energy. Eq. (3-1) can be transformed to
$$
\ln \rho = 2(C/ k_{B})^{1/2}T^{-1/2} + const. 
\eqno(3.2)$$

The linear relation found experimentally in Fig. 3-4 agrees with Eq. (3-2). 
From the gradient of the log $\rho $ versus $T^{-1/2}$, $C$ is estimated to be 
30 meV for the Co$_{36}$Al$_{22}$O$_{42 }$film. 

Figs. 3-5 (a) and (b) show $M-H$ and $M-T$ curves, respectively, for the 
Co$_{36}$Al$_{22}$O$_{42}$ film. The M-H curves at RT show no remanence at 
zero field and unsaturated behavior in high fields. Large thermal hysteresis 
is observed between zero-field cool (ZFC) and field cool (FC) curves, and it 
is suggested that particles are superparamagnetic at room temperature and 
that thermal fluctuation of the magnetic moments is blocked at 4.5 K. Figs. 
3-6 (a) and (b) show MR curves for Co$_{36}$Al$_{22}$O$_{42}$ measured at RT 
and 4.2 K, respectively. Large values of MR exceeding 10 {\%} are observed. 
At RT MR does not completely saturate even at 80 kOe; on the other hand, MR 
nearly saturates above 20 kOe at 4.2 K. The coercivity and hysteresis of the 
MR curve appearing at 4.2 K corresponds to the magnetization curve shown in 
Fig. 3-5 (a). \newline

\textbf{3.3 Particle size distribution in insulating granular systems}\newline

The sizes of nanoparticles in insulating granular films are distributed; 
information on the particle size distribution is essential to understanding 
the mechanism of TMR phenomena [85]. Ohnuma et al. [83, 
86] reported the microstrustures in Co-Al-O granular films investigated by 
high-resolution transmission electron microscopy (HRTEM) to obtain the 
distributions of particle sizes and interparticle distances. We have 
evaluated particle size distributions in Co$_{36}$Al$_{22}$O$_{42}$ granular 
films by fitting the magnetization curves to the Langevin function which 
describes superparamagnetic behavior. 

Magnetization in the superparamagnetic state can be described by the 
Langevin function. If we assume that all the Co particles have a spherical 
body with the same diameter $d$ and if the anisotropy energy is negligible, the 
magnetization $M$ is described by
$$
\frac{M}{M_s }=L\left( \alpha \right)=\coth \alpha -\frac{1}{\alpha }
\eqno(3.3)$$
with
$$
\alpha =\frac{\mu H}{k_B T}=\frac{M_{Co} H}{k_B T}\frac{4\pi }{3}\left( 
{\frac{d}{2}} \right)^3\quad ,
\eqno(3.4)$$
where $M$s is the saturation magnetization of the sample, $M$Co is the 
magnetization of Co particles and $H$ is the applied field. In real systems, 
$d$ has a distribution; we use a log-normal distribution function (LNDF) in 
$d$. LNDF is described by
$$
f\left( d \right)=\frac{1}{\sqrt {2\pi } \ln \sigma }\exp \left[ 
{-\frac{\left( {\ln d-\ln d_m } \right)^2}{2\ln ^2\sigma }} \right]\quad ,
\eqno(3.5)$$
where the parameters $d$m and \textit{$\sigma $} represent the statistical median and the 
geometric standard deviation, respectively. The Langevin function taking the 
log-normal particle size distribution into account is described by
$$
M=\sum\limits_{i=1}^n {\left[ {M_{Co} \frac{4\pi }{3}\left( {\frac{d_i }{2}} 
\right)^3f\left( {d_i } \right)L\left( {\alpha _i } \right)} \right]} \eqno(3.6)
$$
with
$$
\alpha _i =\frac{M_{Co} H}{k_B T}\frac{4\pi }{3}\left( {\frac{d_i }{2}} 
\right)^3\quad .\eqno(3.7)
$$
Fig. 3-7 (a) shows the calculated magnetization curves compared to the 
experiment for a Co$_{36}$Al$_{22}$O$_{42}$ film at $T$ = 200 and 293 K; the 
particle size distribution which gives the best fit is shown in Fig. 3-7 
(b). The sizes of particles are distributed mostly in the range of 1.0 to 
5.0 nm in diameter and $d$mag is estimated to be 2.45 nm. This $d$mag is 
consistent with that evaluated from the TEM image $d_{TEM}$ (= 2.20 nm). 

\newpage 
\textbf{4. Spin-dependent higher order tunneling in Coulomb blockade regime 
}\newline

In this chapter, we report on the temperature and bias-voltage dependence of 
MR in insulating Co-Al-O granular films without performing microfabrication. 
The MR exhibits an anomalous increase at low temperatures but no significant 
indication of change with bias voltage. We show that the anomalous increase 
of the MR indicates evidence for higher-order tunneling (cotunneling) 
between large granules through intervening small granules [36, 62]. 
We emphasize that the existence of higher-order tunneling is a natural 
consequence of the granular structure, since a broad distribution of granule 
size is an intrinsic property of granular systems.\newline

\textbf{4.1 Results}\newline

Fig. 4-1 shows the temperature dependence of MR for Co-Al-O films. It is 
clearly seen that the MR is remarkably enhanced at low temperatures while it 
is nearly independent of temperature above $\sim $100 K. For 
Co$_{36}$Al$_{22}$O$_{42}$, the MR below 3 K is anomalously large and 
reaches more than twice the value given by $P^{2}_{Co}$/(1+$ P^{2}_{Co})$ 
[11], where the formula is half of that for magnetic tunnel junctions 
(MTJ) because of the difference between random and antiparallel alignment of 
magnetic moments. In the case of MTJ with an Al-O barrier, the temperature 
dependence of MR is discussed on the basis of magnetic impurity or magnon 
scattering. However, it is considered that magnetic impurity or magnon 
scattering does not give rise to the plateau in the temperature dependence 
of MR, as in the present result observed above $\sim $100 K. Helman and 
Abeles [80] proposed a theory of spin-dependent tunneling in 
insulating granular systems and predicted the temperature dependence 1/$T$ for 
MR. However, the dependence 1/$T$ does not fit the present results. 

Transport properties were measured in the current-perpendicular-to-plane 
(CPP) geometry as shown in Figs. 4-2 where a 1-$\mu $m-thick 
Co$_{36}$Al$_{22}$O$_{42}$ granular film was sandwiched between upper and 
lower Au-Cr electrodes. Figs. 4-3 (a) and (b) show \textit{$\rho $} and MR at 4.2 K, 
respectively, as functions of bias voltage $V_{b}$ for a 
Co$_{36}$Al$_{22}$O$_{42}$ film. \textit{$\rho $} decreased rapidly by 3 orders of magnitude 
as the bias voltage increased from $V_{b}$ = 0 up to 600 mV. Nevertheless, 
the magnitude of the enhanced MR was almost constant. This is in clear 
contrast with the case of MTJ of macroscopic size, where both MR and \textit{$\rho $} 
decrease gradually with increasing bias voltage. Furthermore, the 
bias-voltage dependence of MR is much smaller than the temperature 
dependence of MR in Fig. 4-1. We can consider that about 200--300 Co 
granules exist in the direction normal to the film plane between the upper 
and the lower electrodes, assuming an average particle size ${\left\langle d \right\rangle}$ of 2.5 nm 
and interparticle distance ${\left\langle s \right\rangle}$ of 1 nm. Therefore, the applied bias 
voltage per one microjunction consisting of two neighboring Co granules may 
be estimated to be 2--3 mV at $V_{b}$ = 600 mV, which corresponds to 20--30 K 
in temperature. As seen in Fig. 4-1, the enhanced MR decreases rapidly with 
increasing temperature and becomes flat around 20--30 K, while it is 
independent of $V_{b}$ at least up to 600 mV (2-3 mV at neighboring ones). \newline

\textbf{4.2 Theoretical explanation}\newline

In granular systems with a broad distribution in particle size [85], 
it is highly probable that large particles are well separated from each 
other due to their low number density (i.e., the larger the granule size, 
the more separated the granules), and there may be a number of smaller 
granules between large granules as shown in Fig. 4-4 (a). To model the 
structural feature of granular systems we assume that large granules with 
size $n{\left\langle d \right\rangle}$ and charging energy ${\left\langle E_{c} \right\rangle}$/$n$ are separated by an array of 
$n$ particles with average size $${\left\langle d \right\rangle}$$ and charging energy ${\left\langle E_{c} \right\rangle}$ on a 
conduction path, as shown in Fig. 4-4 (b).\newline

\textbf{4.2.1 Temperature dependence of MR}

We first calculate the temperature dependence of the conductivity \textit{$\sigma $}($T)$ at 
zero-bias voltage ($V_{b}$ = 0). The tunnel current at the zero-bias limit is 
dominated by thermally activated charge carriers. In the case of the 
conduction path in Fig. 4(b), the carriers mostly occupy the large particle 
of charging energy ${\left\langle E_{c} \right\rangle}$/$n$ in a probability proportional to the 
Boltzmann factor $\exp [-\left\langle {E_c } \right\rangle /2nT]$ in units 
of $k_{B}$ = 1. Since the large particles are separated by the smaller ones, 
the ordinary tunneling of an electron from a large particle to a small one 
increases the charging energy by $\delta E_c \sim \raise0.7ex\hbox{$1$} 
\!\mathord{\left/ {\vphantom {1 
2}}\right.\kern-\nulldelimiterspace}\!\lower0.7ex\hbox{$2$}\left\langle {E_c 
} \right\rangle /(1+1/n)$; thus is suppressed by the Coulomb blockade at low 
temperatures $T \quad < \quad \delta E_{c}$. In this regime, the dominant contribution 
to\textit{${\rm g}\sigma $}($T)$ comes from higher-order processes of spin-dependent tunneling where the 
carrier is transferred from the charged large particle to the neighboring 
neutral large particle through an array of small particles, using successive 
tunneling of single electrons, i.e., the cotunneling of ($n$+1) electrons. 
Figure 4-4 (a) shows an example of the third-order process ($n$=2). Summing up 
all of these higher-order processes, we have
$$
\sigma (T)\propto \sum\limits_n {e^{-\left\langle {E_c } \right\rangle 
/2nT}[(1+P^2m^2)e^{-2\kappa {s}'}]^{n+1}} \left( {\frac{\left( {\pi T} 
\right)^2}{\left( {\delta E_c } \right)^2+\gamma ^2\left( T \right)}} 
\right)^nf\left( n \right) \eqno(4.1)
$$
Here, [---] is the spin-dependent tunneling probability between the 
neighboring particles, $m=M$/$M_{s}$ is the magnetization normalized to the 
saturation magnetization $M_{s}$, \textit{$\kappa $} is the tunneling parameter related to the 
barrier height, and ${s}'=2n\left\langle s \right\rangle /(n+1)$ with 
$\left\langle s \right\rangle $ being the mean separation of particles with 
size$\left\langle d \right\rangle $. The factor (---)$^{n}$ represents the 
finite temperature effect by which electrons (or holes) in the energy 
interval of \textit{$\pi $T} around the Fermi level participate in the intermediate states 
of the higher-order process [87], and \textit{$\gamma $}($T)$ is the decay rate given by 
$\gamma (T)\approx gT$ with $g$ being a constant .[88]. The function 
$f(n)$ represents the distribution of the conduction paths. In Eq. (4.1), $\exp 
\left[ {4\tilde {\kappa }n\left\langle s \right\rangle -\left\langle {E_c } 
\right\rangle /2nT} \right]$ is a peaked function of $n$ and has its maximum at 
$n\ast =\left( {\left\langle {E_c } \right\rangle /8\tilde {\kappa 
}\left\langle s \right\rangle T} \right)^{1/2}$, where $\tilde {\kappa 
}/\kappa \approx 1+\left( {1/4\kappa \left\langle s \right\rangle } 
\right)\ln \left[ {(g/\pi )^2+(\left\langle {E_c } \right\rangle /2\pi T)^2} 
\right]$. The existence of higher-order tunneling processes with different 
orders is a natural consequence of the granular structure since a broad 
distribution of granule sizes is an intrinsic property of granular systems. 
Namely, cotunneling processes with $n$ = 1 occur in some places, while those 
with $n$ = 2 occur in other places, thus $n$ (and consequently $n$*) can be treated 
as a continuous variable at low temperatures $\left( {T<<\left\langle {E_c } 
\right\rangle } \right)$. Replacing the summation by the integration in Eq. 
(4.1) and using the method of steepest descent [89], we obtain
$$
\sigma \left( T \right)\propto \left( {1+P^2m^2} \right)^{n\ast +1}\sqrt 
{\frac{n\ast }{\tilde {\kappa }\left\langle s \right\rangle }} f(n\ast )\exp 
\left[ {-2\sqrt {\frac{2\tilde {\kappa }\left\langle s \right\rangle 
\left\langle {E_c } \right\rangle }{T}} } \right]\quad . \eqno(4.2)
$$
In Fig. 4-5, the calculated resistivity for $m$ = 0 is shown by the solid 
lines. Here and hereafter, we assume $f\left( {n\ast } \right)\propto 
1/n\ast $, and take $2\kappa \left\langle s \right\rangle =3$, g = 0.3, and 
the values of $\left\langle {E_c } \right\rangle $ are estimated to be 9 K 
for Co$_{54}$Al$_{21}$O$_{25}$, 18 K for Co$_{52}$Al$_{20}$O$_{28}$, 25 K 
for Co$_{46}$Al$_{19}$O$_{35}$, and 110 K for Co$_{36}$Al$_{22}$O$_{42}$. 

Because of the higher-order processes, the spin-dependent part of \textit{$\sigma $}($T)$ in Eq. 
(4.2) is amplified to the ($n$* + 1)th power of (1 + $P^{2}m^{2})$, so that 
\textit{$\sigma $}($T)$ is sensitive to the applied magnetic field since $m$ varies from $m$ = 0 to $m$ = 
1 (the fully magnetized state) by application of the magnetic field. Using 
Eq. (2) the MR, $\Delta \rho /\rho _0 =1-\left[ {\sigma \left( T \right)} 
\right]_{m=0} /\sigma \left( T \right)$, is expressed as 
$$
\frac{\Delta \rho }{\rho _0 }=1-\left( {1+m^2P^2} \right)^{-\left( {n\ast +1} \right)}. \eqno(4.3)
$$
The calculated MR is shown by the solid curves in Fig. 4-1, where the value 
of $P$ is chosen to fit the experimental data. For small $P^{2}$, Eq. (3) is 
approximated to be
$$
\frac{\Delta \rho }{\rho _0 }=P^2m^2\left( {1+\sqrt {\frac{C}{T}} } \right) \eqno(4.4)
$$
with $C=\left\langle {E_c } \right\rangle /8\tilde {\kappa }\left\langle s 
\right\rangle $ being constant. Eq. (4.4) indicates an anomalous increase of 
$\Delta \rho /\rho _0 $ at low temperatures due to the onset of higher-order 
processes between larger granules, i.e., $n\ast \propto 1/\sqrt T $. At $T$ = 2 
K, $n$* takes the value of 1.6, so that one or two small granules intervene 
between larger ones in the higher-order processes. As seen in curve a, the 
MR grows rapidly around 10 K well below $E_{c}$ = 110 K. Similar behavior is 
seen in a double-junction system [41] [36]. \newline

\textbf{4.2.2 Bias voltage dependence of MR}

We next calculate the bias-voltage dependence of conductivity \textit{$\sigma $}($V_{b})$ in 
the Coulomb blockade regime. When a finite voltage $V_{b}$ is applied to the 
granular system, the voltage drop $\Delta V_{b}$ between the large granules 
in the model system of Fig. 4-4 (b) is given by $\Delta V_b =\left( {2n/N_g 
} \right)V_b $, where $N_{g}$ is the average number of glanules along a 
conduction path. \textit{$\sigma $}($V_{b})$ at finite temperatures, the factor ($\pi 
T)^{2n}$ in Eq. (4.1), is replaced by $\left[ {\left( {\pi T} 
\right)^2+\left( {2eV_b /N_g } \right)^2} \right]^n$ [87]. We obtain
$$
\sigma (V_b )\propto \sum\limits_n {e^{-\left\langle {E_c } \right\rangle 
/2nT}[(1+P^2m^2)e^{-2\kappa {s}'}]^{n+1}} \left[ {1+\left( {\frac{2eV_b 
}{N_g \pi T}} \right)^2} \right]^nf\left( n \right).\eqno(4.5)
$$
Following the same procedure as in deriving $\sigma (T)$ in Eq. (4.2), we 
obtain the bias --dependence of the conductivity
$$
\sigma \left( {V_b } \right)=\sigma \left( T \right)\left[ {1+\left( 
{\frac{2eV_b }{N_g \pi T}} \right)^2} \right]^{n\ast }.\eqno(4.6)
$$
The \textit{$\sigma $}($V_{b})$ exhibits a power low dependence (1/$V_{b})^{2n\ast }$ for $T\quad<$ 
\textit{eV}$_{b}$/$N_{g}$. 

In Fig. 4-2 (a), we show the calculated resistivity $\rho _0 \left( {V_b } 
\right)=1/\sigma \left( {V_b } \right)$ by the solid curves for $T$ = 4.2 K and 
$N_{g}$ = 140. The steep decrease of the calculated resistivity is in good 
agreement with that of the experimental data. In Fig. 4-3 (b), the 
calculated MR is shown by the solid curve. The enhanced MR is maintained 
upon application of higher voltages, which is consistent with the 
experimental result. The constant MR may originate from the large number of 
granules along the conduction paths in the granular films, in which the 
voltage drop between neighboring granules $\sim V/N_g $ is small for a large 
value of $N_{g}$. $V$ / $N_{g}$ at $V_{b}$ = 500 mV is about 3 mV for $N_{g}$ = 140 
and its corresponding temperature is $\sim $ 40 K, which is smaller than 
that of the charging energy $\left\langle {E_c }\right\rangle \quad \sim$ 110 
K. We note that the enhanced MR is nearly constant up to 500 meV, whereas 
the corresponding resistance is reduced by several orders of magnitude. This 
is in contrast with ferromagnetic tunnel junctions of macroscopic size, 
where both the MR and the resistance decrease gradually with increasing bias 
voltage.

\newpage 
\textbf{5. Spin-dependent single-electron tunneling in microfabricated 
structures}\newline

\textbf{5.1 Current path restriction to observe spin-dependent 
single-electron tunneling in sequential tunneling regime}\newline

Insulating granular films consisting of nanometer-sized magnetic metallic 
particles embedded in an insulating matrix are useful for the study of 
spin-dependent SET (SD-SET) phenomena. The properties of SD-SET are roughly 
divided into two categories according to transport mechanisms: cotunneling 
and sequential tunneling. In the last chapter, we discussed one 
characteristic SD-SET behavior due to cotunneling in the Coulomb blockade 
regime. Here we discuss about the SD-SET behaviors in a sequential tunneling 
regime: in other words, TMR behaviors associated with a Coulomb staircase 
and/or clear Coulomb threshold, in microfabricated samples. 

In granular film of a macroscopic size containing a large number of 
particles, however, SET phenomena, represented by Coulomb staircase and so 
on, are averaged out due to the large distributions of particle sizes and 
interparticle distances [62, 83, 85]. The tunnel paths 
should be restricted in order to clearly observe the SET phenomena. A simple 
method to restrict the tunnel paths is to use scanning tunneling microscopy 
(STM). The tunnel path on the surface is limited to only one particle just 
below the STM tip. We observed clear Coulomb staircases in the $I-V_{b}$ 
measurements for Co-Al-O granular films even at room temperature 
[90, 91]. Figs. 5-1 (a) and (b) show typical examples of an STM 
topographic image and an $I-V_{b}$ curve, respectively, for a 
Co$_{36}$Al$_{22}$O$_{42}$ film.

A more advantageous method than STM for a variety of measurements and 
applications is to fabricate a device structure consisting of a small part 
of granular film with nanofabricated electrodes. In this study we have 
fabricated two types of device structures with Co-Al-O granular films using 
focused ion-beam (FIB) milling or electron-beam lithography techniques. One 
is a granular nanobridge structure [60, 67]: point-shaped 
electrodes separated by a very narrow lateral gap filled with Co-Al-O 
granular film. The other is a current-perpendicular-to-plane (CPP) geometry 
structure [63-65]: a thin Co-Al-O granular film sandwiched 
by ferromagnetic electrodes with the current flowing in the direction 
perpendicular to the film plane through a few Co particles. We measured the 
$I-V_{b}$ curves in these samples, and found the enhancement and oscillation of 
TMR due to spin-dependent SET in a sequential tunneling regime.\newline

\textbf{5.2 Results in nanobridge structures}\newline

\textbf{5.2.1 Experimental procedures and results}

In this section, we report enhanced TMR just above the Coulomb blockade 
threshold in a sequential tunneling regime [60]. We fabricated 
point-shaped electrodes separated by a very narrow lateral gap filled with 
insulating granular film; we call the resulting structures ``granular 
nanobridges''. As mentioned in the last section, the mechanism for TMR 
enhancement is different from that for higher-order tunneling, because this 
enhancement in nanobridge samples is caused outside the Coulomb blockade 
regime. We apply the orthodox theory of SET [9] and explain that 
enhanced TMR is brought about by the modification of the detailed balance of 
particle charges by the external magnetic field [14, 16, 91]. 

A schematic view of a typical sample is shown in Fig. 5-2 (a). An insulating 
granular nanobridge was fabricated on a Corning no. 7059 glass substrate as 
follows: a 15 nm-thick-NbZrSi amorphous layer was deposited by rf 
sputtering, and was formed into source and drain electrodes by FIB etching 
using 30 kV gallium ions (Seiko Instruments Inc., SMI 9200). The electrodes 
separated by a gap with a length ($l)$, i.e., gap separation, of 30 nm and a 
width ($w)$ of 60 nm are shown in Fig. 5-2 (b). Deep trenches (60 nm wide and 
200 nm deep) were formed beside the gap by FIB etching to avoid the 
formation of unnecessary current paths outside the gap. A 7.5-nm-thick 
Co-Al-O granular film was deposited on the patterned surface by reactive rf 
sputtering, and the gap was filled by the Co-Al-O film. The aspect ratio of 
trenches was so high that the trenches were not filled with the Co-Al-O 
film. The composition of Co-Al-O was determined to be 
Co$_{36}$Al$_{22}$O$_{42}$ by RBS analysis. The average size of Co particles 
was estimated to be about 2.5 nm from the analysis of the superparamagnetic 
behavior and TEM observation [85]. The characteristic sizes of 
granular nanobridges, i.e., $w$, $l$, and thickness ($t)$, varied in the range of 60 
- 700 nm, 30 - 70 nm and 5 - 30 nm, respectively. $I-V_{b}$ characteristics 
were measured at 4.2 K using an electrometer (Keithley 6514) with a 
two-terminal arrangement. TMR (= $\Delta R/ R_{max})$ was evaluated from the 
difference between the $I-V_{b}$ curves at the applied field $H$ = 0 and 10 kOe.

Fig. 5-3 (a) shows the $I-V_{b}$ curves at $H$=0 (solid lines) and $H$=10 kOe (dashed 
lines) for the sample with $w$=60 nm, $l$=30 nm and $t$=7.5 nm. Here, the threshold 
voltage $V_{th}$ was defined as that below which the current was zero within 
an accuracy of 100 fA, and it was approximately 1.5 V in this case. In the 
range of $\left| {V_b } \right|<V_{th}$, a Coulomb blockade occurred, and 
the current increased rapidly when $\left| {V_b } \right|$ exceeded 
$V_{th}$. It is noted that for Co-Al-O granular films of macroscopic sizes, 
the Coulomb blockade has not clearly been observed, because the macroscopic 
sample contains a lot of Co particles with a broad distribution of sizes and 
the tunneling of electrons between large particles with small $E_{c}$ could 
start at small voltages. In the granular nanobridge, however, the tunnel 
paths are so limited that the Coulomb blockade is remarkable.

Fig. 5-3 (b) shows the $V_{b}$ dependence of TMR. TMR depends strongly on 
$V_{b}$. For $\left| {V_b } \right|<$4.0 V, TMR increases with decreasing 
$\left| {V_b } \right|$ and reaches a maximum value larger than 30 {\%} at 
the voltage slightly above $V_{th}$ ($\sim $1.5 V). In the Coulomb blockade 
region, $i. e$., $\left| {V_b } \right|<V_{th}$ (hatching area), there is 
little quantitative reliability of the measurements because the current is 
very low ($<$100 fA). For $\left| {V_b } \right|>$ 4.0 V, on the other 
hand, TMR is about 8 {\%} showing no large change with $V_{b}$, although weak 
oscillatory behavior seems to exist in the $V_{b}$ dependence of TMR.

Similar results have been obtained in other granular nanobridges of 
different sizes. Figs. 5-4 (a) and (b) show the $I-V_{b}$ curve and the 
$V_{b}$ dependence of TMR, respectively, for the sample with $w$=700 nm, $l$=40 nm 
and $t$=15 nm. $V_{th}$ is observed to be 0.5 V, which is lower than that in the 
sample shown in Fig. 5-3. $V_{th}$ shows a tendency to increase as the size 
of the granular nanobridge decreases. The voltage where the TMR shows a 
maximum, $V_{p}$, is slightly larger than $V_{th}$. Fig. 5-5 shows $V_{p}$ vs. 
$V_{th}$ in granular nanobridges of different sizes. A clear correlation 
between $V_{p}$ and $V_{th}$ is seen, suggesting that the enhanced TMR is 
caused by the Coulomb blockade.\newline

\textbf{5.2.2 Theoretical explanation}

The orthodox theory of SET [9] can be used to explain the experimental 
results, particularly the enhanced TMR near $V_{th}$. In the orthodox theory, 
the tunnel path is modeled as an equivalent classical electrical circuit. 
For the granular nanobridge, we assume a parallel circuit of triple tunnel 
junctions as shown in Fig. 5-6 (a). This is the simplest model to explain 
our experimental results because we need at least two magnetic particles in 
each series of junctions to study the spin-dependent transport in 
nanobridges with nonmagnetic electrodes. We neglect the higher-order 
tunneling process, i.e., cotunneling [41, 62], because the tunnel 
resistances between particles and between an electrode and a particle are 
estimated to be about 10$^{5}$ times larger than $R_{Q}\approx $25.8 
k$\Omega $. In order to obtain a stable tunneling current, we constructed a 
detailed balance equation for the probability of states $p(\{n_i \}_\alpha 
)$, which is given in the matrix form by
$$
\dot {p}={\rm {\bf M}}p=0,\eqno(5.1)
$$
where $p=(\ldots ,p(\{n_i \}_\alpha ),\ldots )^T$, and \textbf{M} is the 
transition matrix in the configuration space constructed by $\{n_i \}_\alpha 
$ with the index \textit{$\alpha $} labeling the different charge states. The tunneling 
current through the $k^{th}$ junction is given by
$$
I_k =e\sum\nolimits_\alpha {p(\{n_i \}_\alpha )\left[ {\Gamma _k^+ (\{n_i 
\}_\alpha )-\Gamma _k^- (\{n_i \}_\alpha )} \right]} ,\eqno(5.2)
$$
where $\Gamma _k^{+(-)} (\{n_i \}_\alpha )\propto 1/R_k $ is the forward 
(backward) tunneling rate through the $k^{th}$ junction with the initial 
charge state $\{n_i \}_\alpha $. Current conservation requires that the 
tunneling current $I_k $ is the same for all the junctions in each series. 
Let us evaluate the tunneling current at the junction between a nonmagnetic 
electrode and a particle, where the tunneling rate $\Gamma _k^{+(-)} (\{n_i 
\}_\alpha )$ is independent of the magnetic field. The magnetic field 
dependence of the tunneling current comes from the probability $p(\{n_i 
\}_\alpha )$ of the charge state $\{n_i \}_\alpha $, which is determined by 
Eq.(5.2). Since the transition matrix \textbf{M} contains the tunneling 
rates between magnetic particles, the matrix \textbf{M} and therefore the 
probability $p(\{n_i \}_\alpha )$ can be modified by applying the magnetic 
field. For $V_{b}$ just above $V_{th}$, we have a few charge states 
contributing to the tunneling current, and the tunneling rates at these 
charge states differ greatly from each other according to thier charging 
energies. Therefore, the strong modification of the probability $p(\{n_i 
\}_\alpha )$ is made to satisfy the detailed balance equation, and the TMR 
is strongly enhanced just above $V_{th}$. This kind of TMR enhancement in 
double tunnel junctions has been studied by Barnas and Fert [14] and by 
Majumdar and Hershfield [16]. They also predicted the oscillating 
behavior of TMR against $V_{b}$ associated with the Coulomb staircase. No 
Coulomb staircase appears in the $I-V_{b}$ curves of granular nanobridges; 
however, a small-magnitude oscillation of TMR could be observed (see Figs. 
5-3 and 5-4). There is no strong asymmetry in tunnel resistances, i.e., no 
bottleneck in tunnel paths of granular nanobridges, leading to no Coulomb 
staircase and only weak oscillation of TMR. Moreover, we have many junction 
arrays in a granular nanobridge, and the randomness of junction capacitances 
also muddies the TMR oscillation.

In order to explain the experimental results for the sample with $w$=60 nm, 
$l$=30 nm and $t$=7.5 nm (Fig. 5-3), we considered a parallel circuit of 20 triple 
tunnel junctions and assumed that the tunnel resistance between an electrode 
and a particle is expressed as $R_{ep} =(1\pm \delta )\overline R _{ep} $, 
where \textit{$\delta $} is the deviation from the typical value $\overline R _{ep} $. Other 
junction parameters such as tunnel resistances between particles $R_{pp}$, 
junction capacitances $C_{ep}$, and $C_{pp}$ were also assumed to be 
distributed around the mean values, $i.e.$, $R_{pp} =(1\pm \delta )\overline R 
_{pp} $, $C_{ep} =(1\pm \delta )\overline C _{ep} $, and $C_{pp} =(1\pm 
\delta )\overline C _{pp} $. The deviation \textit{$\delta $} for each junction parameter was 
randomly chosen within the range of -0.1$<$\textit{$\delta $}$<$0.1. The temperature was set 
to be 4.2 K and the typical value of tunnel resistances for the parallel 
alignment of magnetizations was taken to be
$$
\overline R _{pp} =\overline R _{ep} /2.\eqno(5.3)
$$
The tunnel resistance between particles for the antiparallel alignment of 
magnetizations was larger than that for the parallel alignment and is 
expressed using the spin polarization $P$ as
$$
\overline R _{pp} =(\overline R _{ep} /2)\cdot (1+P^2)/(1-P^2),\eqno(5.4)
$$
where $P$ is assumed to be 0.42 for Co [92]. The typical 
values of junction capacitances are taken to be $\overline C _{ep} 
=0.1{aF}$ and $\overline C _{pp} =0.05 {aF}$, which are reasonable 
values considering the average particle size and interparticle distance in 
Co-Al-O granular films [83, 85].

The $V_{b}$ dependence of TMR obtained by the numerical calculation is shown 
in Fig. 5-6 (b). One can see that the theoretical result is in good 
agreement with the experimental one. The TMR is enhanced just above 
$V_{th}$ ($\sim $1.5 V) and decreases with $V_{b}$. The randomness of junction 
capacitances obscures the oscillation of the total TMR as shown in Fig. 5-6 
(b). We may consider that the quantitative difference between $V_{th}$ and 
$V_{p}$ is not essential, but it is caused by effects such as the leakage of 
current through the glass substrate.\newline

\textbf{5.3 Results in CPP geometry structures}\newline

In the last section, we fabricated nanobridge structures for combining 
insulating granular films with microfabricated electrodes and successfully 
found enhanced TMR at the Coulomb threshold voltage. Proper limitation of 
the number of current paths made it possible to observe spin-dependent SET. 
However, no Coulomb staircase was observed. In an assembly of nanoparticles 
such as granular films, the Coulomb staircase is expected to appear when the 
tunnel resistance between two neighboring particles or between a particle 
and an electrode is much larger than the other resistances in the current 
path. In other words, a bottleneck of tunnel conductance must exist 
somewhere in the current path [9, 36]. In this section, in order to 
investigate the relationship between the Coulomb staircase and TMR, we 
employed CPP (Current-Perpendicular-to-Plane) geometry measurements in 
Co-Al-O granular films. A bottlenech is easily added to samples for CPP 
geometry measurements for the sake of observing Coulomb staircases. We 
prepared CPP geometry samples, where a thin Co-Al-O granular film was 
sandwiched by ferromagnetic electrodes and a very thin Al-O layer was 
inserted between the bottom electrode and Co-Al-O as a bottleneck, and 
measured the current ($I)$ -- bias voltage ($V_{b})$ characteristics. We 
succeeded in observing clear Coulomb staircases due to the current 
confinement among the vast number of channels between the upper and lower 
electrodes [65].

We fabricated CPP geometry samples by a focused ion-beam (FIB) etching 
process. Fig. 5-7 (a) represents a schematic illustration of a sample. 
Samples were prepared on Si/SiO$_{2}$ substrates by rf sputtering. A bottom 
electrode was first deposited, and then a thick Al-O film (40 nm) was 
deposited onto the bottom electrode as an insulating layer using an 
Al$_{2}$O$_{3}$ target. Next, a small contact area was made by FIB milling 
(Fig. 5-7 (b)). A contact area of about 0.5 x 0.5 $\mu $m$^{2}$ was 
estimated from the scanning ion microscopy image. The milling process was 
carefully performed to leave a very thin Al-O layer, which contributes to 
forming a bottleneck. After making the contact window, furthermore, a 1$\sim 
$2 nm thick Al-O layer was deposited. Consequently, the bottleneck is given 
by the combination of the residual Al-O and the deposited Al-O layer. A 
7$\sim $12 nm-thick Co-Al-O granular layer followed by a top electrode was 
finally deposited. The deposition of Co-Al-O granular films was done through 
the use of reactive sputtering in a mixture of Ar+O$_{2}$ atmosphere. 
$I-V_{b}$ characteristics were measured at 4.2 K using an electrometer 
(Keithley 6514) with a two-terminal arrangement. TMR (= $\Delta 
R$/$R_{max})$ was evaluated from the difference between the $I-V_{b}$ curves at 
the applied field $H$ = 0 and 10 kOe. The external magnetic field was usually 
applied in the direction parallel to the plane.

Fig. 5-8 (a) shows $I-V_{b}$ curves at $H$ = 0 (solid lines) and $H$ = 10 kOe 
(dashed lines) at 4.2 K. Clear Coulomb staircases are observed for both 
$I-V_{b}$ curves. The first three steps from zero bias appear every 20 mV; 
however, the steps at higher bias voltages do not maintain a regular period. 
In spite of more than 10$^{4}$ parallel current paths between the electrodes 
in the contact area (0.5 x 0.5 $\mu $m$^{2})$, which is much larger than the 
average Co particle size (2.5 nm), clear Coulomb staircases were observed. 
This suggests that the current at low bias voltages preferentially flow 
along certain restricted local paths where the total charging energy 
determined from the sum of capacitances through the path is the lowest. 
Similar results were previously reported in CPP measurements for nonmagnetic 
granular films [93]. It is considered that at higher voltages the 
contribution of various current paths including those with higher charging 
energy appears to lead to the irregular period of Coulomb staircase. 

Fig. 5-8 (b) shows the bias voltage dependence of the TMR derived from the 
two $I-V_{b}$ curves shown in Fig. 5-8 (a). We observed the oscillation of the 
TMR as a function of bias voltage. The peak of TMR repeats with the period 
of the Coulomb staircase. The modification of the $I-V_{b}$ curve by the 
applied field seems to bring about the enhancement of the TMR with the steps 
of the Coulomb staircase, resulting in the oscillation of the TMR. The 
largest TMR (about 20 {\%}) was seen at $V_{b}$ = 15 mV; however, TMR 
converged to almost zero as the bias voltage was further increased. This is 
probably because the barrier quality of the Al-O bottleneck layer is still 
poor, leading to the rapid decrease in TMR with bias voltage. It is also 
noted that a sign change was observed in the TMR. One possible origin we may 
consider is the effect of spin accumulation. In order to clarify TMR 
behavior, we performed a further study considering the spin accumulation 
effect in a nanoparticle.

\newpage 
\textbf{6. Spin accumulation effect in nanoparticles}\newline

We observed TMR oscillation associated with the Coulomb staircase in CPP 
geometry samples fabricated by FIB. In this chapter, we report experimental 
evidence for the spin accumulation effect in Co nanoparticles leading to the 
oscillation of TMR with alternate sign changes. Furthermore, the spin 
relaxation time in Co nanoparticles was evaluated by comparing the 
experimental results to numerical simulations. We discovered that the spin 
relaxation time in the nanoparticles is unprecedentedly enhanced up to more 
than the order of hundreds of nano seconds, compared to that evaluated from 
the spin diffusion length of ferromagnetic layers in previous CPP-GMR 
studies, $i.e., $the order of tens of picoseconds. \newline

\textbf{6.1 Experimental procedures and results}\newline

In order to study this subject, the sample design and the fabrication 
process were modified. We fabricated pillared structures consisting of Al 
electrode / Al-O / Co-Al-O granular film / Co electrode layers (Fig. 6-1) to 
measure their magnetotransport properties in CPP geometry. Samples were 
prepared on thermally oxidized silicon substrates as follows: the bottom Al 
electrode was first deposited using ion beam sputtering (IBS), and the 
surface was plasma oxidized to form a conductance bottleneck layer for the 
observation of Coulomb staircase. A 15 nm thick Co-Al-O granular film was 
then deposited by reactive rf sputtering. The top Co electrode and Pt layer 
were deposited by IBS. This layered structure was then microfabricated to 
reduce the contact area using electron-beam (EB) lithography and an Ar ion 
etching process. An Al / Al-O / Co-Al-O / Co / Pt pillar 0.4 $\mu $m x 0.4 
$\mu $m in area was prepared by the following process: EB patterning on 
positive resists, depositing of etching mask (Co), lift-off and Ar ion 
etching. To prevent short-circuiting between the electrodes, the sidewall of 
the pillar was then plasma-oxidized and covered with a thick Al-O also 
deposited by IBS. Finally a Pt layer was deposited to contact the top 
electrode. The sample structure is schematically shown in Fig. 6-1. 

Fig. 6-2 (a) shows the $I-V $characteristics at $H $= 0 (black line) and $H $= 10 kOe 
(gray line) in the positive bias voltage region. The $I-V $curves reveal a 
definite Coulomb staircase, which was also observed in the negative bias 
voltage region (not shown here). The first step of the Coulomb staircase 
appears at about 15 mV, and the subsequent steps appear at 50 mV, 85 mV and 
120 mV, respectively. We think the current preferentially flowed through a 
single or a few restricted local paths where only one particle was located 
between electrodes for the following reason: When the nominal thickness of 
the Co-Al-O layer was smaller than 15 nm, the samples show neither Coulomb 
blockade nor Coulomb staircase. In other words, 15 nm is the minimum 
thickness to observe the single-electron phenomena. This suggests that the 
Co-Al-O layer has considerable thickness fluctuation, and therefore there 
are some thin parts where only one particle exists in the direction normal 
to the film plane. (If the thickness of the Co-Al-O layer is completely 
uniform, three or four particles should exist in the direction normal to the 
film plane everywhere, and the critical thickness for SET should be much 
smaller than 15 nm because the particle size is a few nm.) Furthermore, the 
Al-O bottleneck layer also has thickness fluctuation, playing a role as a 
path restriction filter. The shape of the staircase measured at $H $= 0 is 
different from that of at $H $= 10 kOe. Comparing the shapes of these two 
curves, it appears that the steps of the staircase at $H $= 0 are steeper than 
those at $H $= 10 kOe, leading to a periodic intersection of the two curves 
around the step points of the staircase. 

Fig. 6-2 (b) shows the $V $dependence of TMR derived from the $I-V $curves shown in 
Fig. 6-2 (a). TMR oscillates with the same period as the staircase, ranges 
from -10 {\%} to +15 {\%} and shows alternate sign changes: negative peaks, 
$i.e., $inverse TMR, appear around the step points of the staircase, whereas 
positive bumps appear between these negative peaks. In a conventional 
ferromagnetic tunnel junction of macroscopic size, TMR should be limited 
within that expected from the spin-polarization factors of ferromagnetic 
components; and it decreases monotonically with bias voltage. The expected 
value of the TMR in this sample should be less than 2{\%} because the 
resistance of the bottleneck which is located at the \textit{spin}-\textit{independent }junction between the 
Al electrode and a Co particle is 10 times larger than that of the other 
\textit{spin}-\textit{dependent }junction. Nevertheless, TMR ranging from -10 {\%} to +15 {\%} was observed, 
indicating a significant enhancement and anomalous oscillation with sign 
change of TMR. In order to confirm the sign of TMR, the magnetic field 
dependence of the electrical resistance (MR curves) was measured. Fig. 6-2 
(c) shows MR curves at $V $= 0.05 V and 0.12 V at which TMR shows negative and 
positive values, respectively. Both MR curves show a hysteretic behavior 
with two peaks according to the coercive force of the Co nanoparticles. The 
MR curve at $V $= 0.05 V shows an increase of electrical resistance ($R)$ with 
increasing applied magnetic field; $i.e., $inverse TMR is observed, whereas the MR 
curve at $V $= 0.12 V shows a decrease of $R $with the magnetic field. It is noted 
that the shift of MR curves toward the positive direction of the magnetic 
field is seen. A possible origin for the shift of the MR curves is the 
exchange bias between Co and CoO formed by the surface oxidization of the Co 
nanoparticles.\newline

\textbf{6.2 Numerical analysis}\newline

We performed numerical calculations on the basis of the orthodox theory with 
an arbitrary spin relaxation time in ferromagnetic tunnel junctions and 
compared it with the $V $dependence of TMR in the present experiment. Since in 
the conducting region the Co-Al-O film contains only one nanoparticle 
between electrodes as mentioned before, we employed a PM / I / FM particle / 
I / FM double-tunnel-junction model (see Fig. 6-3 (a)) to analyze the 
experimental data. We considered the collinear alignment of the 
magnetizations. Although Fig. 6-2 (c) shows that the angle between the 
magnetizations of the Co particle and the Co electrode is large at zero 
magnetic field, it is hard to conclude that the magnetizations were aligned 
to be anti-parallel since the easy magnetization directions of nanoparticles 
were distributed randomly and the direction of the applied magnetic field 
was fixed to a certain direction. If the magnetizations are non-collinearly 
aligned the tunneling resistance for each spin depends on the relative angle 
of magnetizations [11, 94] and we have to calculate the nonequilibrium 
spin distribution matrix [95, 96]. However, it is important to 
note that the calculations for collinear configurations give the lower limit 
of the spin relaxation time. Hereafter, we consider only the collinear 
configurations. 

Without losing generality we can take the spin quantization axis so that the 
minority and majority spins of the nanoparticle are represented by 
$\downarrow $ and $\uparrow $, respectively. The spin accumulation occurs in 
the ferromagnetic island where there is a significant difference in the 
densities of states for minority ($D_\downarrow )$ and majority ($D_\uparrow 
)$ spins at the Fermi level due to the exchange splitting in the $d$-band. 
According to the band calculation of Co, the net spin polarization 
$P_{DOS} \equiv (D_\uparrow -D_\downarrow )/(D_\uparrow +D_\downarrow 
)$ is taken to be -0.73. In addition to $P_{DOS} $, another spin 
polarization must be taken into account, $i.e.$, that for tunneling electrons 
which is dominated by $s-d $hybridization [97]. It is described as 
$P_{tun} $, hereafter. $P_{tun} $of Co is assumed to be 0.35 
[98]. We suppose that the energy relaxation time is so short that the 
Fermi distribution function can be used to represent the electron 
distribution in the Co nanoparticle. The model shown schematically in Fig. 
6-3 (a) assumes that the magnetization vector of the nanoparticle is in a 
fixed orientation, whereas that of the right electrode is reversed by 
applying a magnetic field, leading to parallel (P) and anti-parallel (AP) 
alignments. The TMR is defined as ${\mbox{TMR}}\equiv 1-R_{P}
/R_{AP} $, where $R_P(AP) $ is the resistance of the whole 
system in the P (AP) alignment. The tunnel resistance and capacitance of 
each junction have been adjusted to reproduce the magnitude of the tunneling 
current and the period of the Coulomb staircase. The capacitances are taken 
to be $C_{1}$ = 4.44 aF and $C_{2}$ = 3.00 aF, where subscripts 1 and 2 
correspond to the left and right junctions, respectively. The tunnel 
resistances for the parallel alignment were determined in the following way: 
first we decided that the tunnel resistance of the majority spin for 
parallel alignment at the left junction ($R_{1,\uparrow }^{P} )$ was 
25.0 G$\Omega $, and the conductance bottleneck was placed at the left 
junction (PM / I / FM particle), assuming that $R_{1,\uparrow }^{P} $ 
was 10 times larger than $R_{2,\uparrow }^{P} $. The other resistances 
are given by
$$
R_{1,\downarrow }^{P} =R_{1,\downarrow }^{P} (1+P_{Al})(1+P_{Partcle} )/(1-P_{Al} )(1-P_{Partcle} )\quad=\quad 51.9 G\Omega,  \\ $$
$$R_{2,\uparrow }^{P} =0.1\cdot R_{1,\uparrow }^{P} \quad=\quad250 M\Omega , \\ $$
$$R_{2,\downarrow }^{P} =0.1\cdot R_{1,\uparrow }^{P} 
(1+P_{Particle} )(1+P_{Co} )/(1-P_{Particle} )(1-P_{Co} )\quad=\quad1.08 G\Omega , \eqno(6.1)
$$
where $P_{Co}$, $P_{Particle}$ and $P_{Al}$ are the $P_{tun}$ of the right 
electrode, the particle and the left electrode, and are assumed to be 0.35, 
0.35 and 0, respectively. These considerably large resistances (G$\Omega )$ 
are reasonable for such a small tunnel junction consisting of nanoparticles. 
For the antiparallel alignment, tunnel resistances at the left junction were 
$R_{1,\uparrow }^{AP} =R_{1,\uparrow }^{P} $ = 25.0 G$\Omega 
$ and $R_{1,\downarrow }^{AP} =R_{1,\downarrow }^{P} $ = 51.9 
G$\Omega $. At the right junction, $D_{Particle,\downarrow } 
=D_{Particle,\uparrow } \left( {1-P_{Partcile} } \right)/\left( 
{1+P_{Particle} } \right)$, $D_{Co,\downarrow } =D_{Co,\uparrow } \left( 
{1-P_{Co} } \right)/\left( {1+P_{Co} } \right)$ and $R_{2,\uparrow 
}^{P} \propto (D_{Particle,\uparrow } D_{Co,\uparrow } )^{-1}$produced 
the following resistances:
$$
R_{2,\uparrow }^{AP} =R_{2,\uparrow }^{P} (1+P_{Co} )/(1-P_{Co} )\quad =\quad 519 M\Omega ,$$
$$R_{2,\downarrow }^{AP} =R_{2,\uparrow }^{P} (1+P_{Particle})/(1-P_{Particle} )\quad =\quad 519 M\Omega.\eqno(6.2)
$$
When an external voltage is applied, electrons flow from the left to the 
right electrode. The current flowing in the junction is calculated using the 
master equation technique [21, 22]. By applying the charge conservation and 
Kirchhoff's law and by using the Fermi golden rule for the calculation of 
the tunneling Hamiltonian, we could write the tunneling rate for the $i$-th 
junction as follows:
$$
\Gamma _{i,\sigma }^\pm (n)=\frac{1}{e^2R_{i,\sigma } }\frac{E_i^\pm 
(n)+\Delta E_F^\sigma }{\exp \left( {E_i^\pm (n)+\Delta E_F^\sigma /k_B T} 
\right)-1} \eqno(6.3)
$$
where $i$ stands for the considered junction ($i $= 1, 2), $\sigma $ stands for the 
direction of the spin, and the exponent $\pm $ indicates whether an electron 
is added or removed from the particle. Finally, the energy is written:

$$E_i^\pm (n)=(1+2n)\frac{e^2}{2C}\pm \frac{C_j }{C}eV (where j\ne i)  \eqno(6.4) $$
where $C=C_{1}+C_{2}$, $T$ is the temperature, and $\Delta E_F^\sigma $ the 
shift of the Fermi energy of the island due to spin accumulation for the 
\textit{$\sigma $}-spin population. In the stationary regime, the transition rates for 
incoming and outgoing electrons cancel each other thes and the probability 
p($n)$ of finding $n$ additional electrons in the central particle may then be 
calculated. Finally, the current is expressed as follows:
$$
I=-e\sum\limits_\sigma {\sum\limits_{n=-\infty }^\infty {\left( {\Gamma _{1,\sigma }^+ (n)-\Gamma _{1,\sigma }^- (n)} \right)} } \eqno(6.5)
$$
The spin accumulation in a Co nanoparticle is represented by the 
spin-dependent shift of the chemical potential for the${\rm g}\sigma 
{\rm g}$pin state $\Delta E_F^\sigma $, which is determined by 
charge-neutrality and spin-conservation conditions. The former is given by 
$D_\uparrow \Delta E_F^\uparrow =-D_\downarrow \Delta E_F^\downarrow , $where 
$D_\sigma $ is the density of states at the Fermi level for spin $\sigma $. 
The latter is expressed as follows:
$$
\frac{(I_{1,\sigma } -I_{2,\sigma } )}{e}=\frac{D_\sigma \Omega }{\tau 
_{SF} }\Delta E_F^\sigma \eqno(6.6)
$$
where $I_{i,\sigma } $ is the current at the $i^{th}$ junction ($i $= 1, 2) for 
spin $\sigma $, $\Omega $ is the volume of the particle, $\tau _{SF} 
$ is the mean spin relaxation time in a nanoparticle defined as $\tau 
_{SF}^{-1} =(\tau _\uparrow ^{-1} +\tau _\downarrow ^{-1} )/2$, and 
$\tau _\sigma $ is the spin relaxation time of electrons with spin $\sigma $. 
For a ferromagnet, $\tau _\uparrow $ and $\tau _\downarrow $ are not the same 
but satisfy the detailed balancing equation, $D_\uparrow /\tau _\uparrow 
=D_\downarrow /\tau _\downarrow $. The tunnel current is obtained by solving 
the master equation, charge-neutrality and spin-conservation conditions, 
self-consistently [14, 15, 27, 36]. 

We first show the results of numerical calculation assuming an infinite spin 
relaxation time in Fig. 6-3 (b), (c) and (d). The significant effect of the 
spin accumulation is found in these results. Fig. 6-3 (b) shows the $V $dependence 
of the chemical potential shifts $\Delta E_F^\sigma $ for the majority and 
minority spins in the P and AP alignment. It shows a sawlike oscillation 
with a period close to that of the Coulomb staircase shown in Fig. 6-3 (c). 
The important point is that the difference in the energy shift $\Delta 
E_F^\sigma $ for each alignment and spin leads to a splitting of the 
discrete charging levels $E_c^{\sigma ,P(AP)} (n)$ for spin $\sigma $ 
and P(AP) alignments, where $n $denotes the numbers of the charge in the 
particle. $E_c^{\sigma ,P(AP)} (n)$ is expressed as
$$
E_c^{\sigma ,{P(AP)}} (n)=\frac{(n+1/2)e^2}{C_1 }+\Delta E_F^{\sigma 
,{P(AP)}} (V).\eqno(6.7)
$$
The voltage corresponding to the step points of the Coulomb staircase is 
then accordingly split. In Fig. 6-3 (c), one can see that each $I-V $curve has a 
unique shape because $\Delta E_F^\sigma $ affects the tunnel conductance, and 
furthermore intersects around the step points of the staircase due to the 
split of $E_c^{\sigma ,P(AP)} (n)$. This in turn leads to the 
periodical sign change of the $V $dependence of TMR (TMR-$V $curve) with a period 
close to that of the Coulomb staircase as shown in Fig. 6-3 (d). 

We also performed a calculation assuming an infinite spin relaxation time in 
the case of the converse arrangement of the conductance bottleneck; i.e., 
$R_{2,\uparrow }^{P} $ was fixed to be 25.0 G$\Omega $ and was assumed 
to be 10 times larger than $R_{1,\uparrow }^{P} $. The other 
resistances and capacitances are described in Fig. 6-4 (a). We show the 
results in Fig. 6-4 (b), (c) and (d). Though the $V$ dependence of the chemical 
potential also appears as sawlike oscillations with a period of the Coulomb 
staircase, none of the $I-V$ curve intersect; thus no sign changes occur for TMR 
in the entire $V$ range. These results indicate that the alternate sign change 
of TMR due to spin accumulation in the Co nanoparticle is reproduced when 
the conductance bottleneck is put at the junction of the PM / I / FM 
particle.\newline

\textbf{6.3 Evaluation of spin relaxation time}\newline

Fig. 6-5 shows the TMR-$V $curve for various values of the spin relaxation time 
$\tau _{SF} $. Although the period of the oscillation does not depend 
on $\tau _{SF} $, the shape of the TMR-$V $curve is quite different 
between the fast and slow spin relaxation regimes. The critical value of the 
spin relaxation time which divides the fast- and slow-spin relaxation 
regimes is about 100 ns, which corresponds to the interval between two 
successive tunneling events ($e$/\textit{I $\approx $ }100 ns). In the fast-spin relaxation regime 
($\tau _{SF}  \quad <$ 100 ns), the TMR-$V $curve shows sharp positive peaks 
at the step points of the Coulomb staircase. On the contrary, in the slow 
spin relaxation regime ($\tau _{SF}  \quad >$ 100 ns), sharp \textit{negative }peaks appear 
at the same step points of the staircase, and broad positive bumps appear 
between two successive negative peaks. In Fig. 6-6 (a), (b), (c) and (d), we 
show cross-sections of Fig. 6-5 at $\tau _{SF} $ = 1 ns, 10 ns, 150 
ns and infinity, respectively. At $\tau _{SF} $ = 1 ns, in the 
fast-spin relaxation regime, we have sharp positive peaks at $V $= 18$, $53$, $88, and 
123 mV, where the tunneling current shows steps. The TMR-$V $curve maintains the 
same shape as far as the spin relaxation time $\tau _{SF}  \quad <$ 1 ns. 
However, at $\tau _{SF} $ = 10 ns, one can see that dips which take 
negative TMR values appear at the step points of the Coulomb staircase. At 
$\tau _{SF} $ = 150 ns, the TMR shows alternate sign changes with 
moderate amplitude reaching a maximum at infinite $\tau _{SF} $. The 
negative peaks associated with the steps of the Coulomb staircase provide 
clear evidence of spin accumulation in Co nanoparticles [15, 27]. 
Finally as shown in Fig. 6-7, the experimentally observed TMR curve is well 
reproduced by choosing the spin relaxation time to be $\tau _{SF} $ = 
150 ns. 

The present TMR experiments have shown that the tunnel current through 
nanoparticles is drastically affected by spin accumulation, and have 
revealed that the TMR exhibits a sign change at the step points of the 
Coulomb staircase. Although the spin relaxation mechanism of the magnetic 
nanoparticle is not well understood at present, there are some candidates 
for the cause of such a dramatic increase of the spin relaxation time. One 
is the quantization of energy levels, which is the most characteristic 
feature of nanoparticles with diameters of 1$\sim $5nm 
[99-102]. The discrete energy level spacing in Co 
nanoparticles having a mean diameter of $\sim $2.5 nm is estimated to be 
about 2 meV, which is larger than the thermal fluctuation at 4.2 K ($\sim 
$0.4 meV). The discreteness of the energy levels due to the zero 
dimensionality of nanoparticles leads to strong suppression of spin-flip 
scattering by the spin-orbit interaction in the nanoparticles [103, 
104]. Mitrikas \textit{et al.} have studied the spin-lattice relaxation of paramagnetic 
nanoparticles embedded in amorphous (SiO$_{2})$ and a crystalline 
(TiO$_{2})$ matrices [105]. They showed that the spin lattice 
relaxation (SLR) is blocked due to the amorphous phase of the matrix. 
Although their nanoparticle is not ferromagnetic, the blocking of the SLR is 
another candidate for the strong enhancement of the spin relaxation time in 
our sample because the insulating Al-O matrix is in an amorphous phase. 
Furthermore, the spin-flip mechanisms due to the interaction with magnons in 
a nanoparticle are also suppressed because of the size quantization of 
magnon excitations [106]. 

Recently, further experimental studies on spin accumulation in paramagnetic 
Au nanoparticles have been performed [68, 72]. Observation 
of non-zero TMR is attributed to the spin accumulation in Au nanoparticles 
in an FM / I / Au particles / I / FM structure, which should not show TMR in 
the framework of Julli\'{e}re model. Although these studies did not mention 
the spin relaxation time in Au nanoparticles, the nano-ampere order of the 
observed current suggested a nano-second order for the spin relaxation time 
($e$/$I)$. It is suggested to be enhanced even in the Au nanoparticles. All of 
these particular features of nanoparticles can lead to an extremely long 
spin relaxation time and a striking spin accumulation effect on TMR. To 
determine the dominant mechanism of spin relaxation in nanoparticles, 
further studies investigating the dependence of spin relaxation time on 
variables such as the temperature, the material of the insulating matrix and 
particle size are needed. We note that the results shown here are promising 
for the potential application of nanoparticles as basic elements of 
spin-electronic devices.

\newpage 
\textbf{7. Summary}\newline

We have reviewed studies on spin-dependent transport in systems containing 
ferromagnetic nanoparticles. In a tunnel junction with a ferromagnetic 
nano-island and electrode, spin-dependent single-electron tunneling (SD-SET) 
gives rise to remarkable tunnel magnetoresistance (TMR) phenomena. We 
studied magnetotransport properties in both sequential tunneling and 
cotunneling regimes of SET and found the enhancement and oscillation of TMR. 

The self-assembled ferromagnetic nanoparticles we used in this study 
consisted of a Co-Al-O granular film, with cobalt nanoparticles embedded in 
the Al-O insulating matrix. In Ch. 3 we showed the preparation methods, and 
basic transport and magnetic properties of the films. A 
Co$_{36}$Al$_{22}$O$_{42}$ film prepared by the reactive sputtering method 
produced a TMR ratio reaching 10 {\%} and superparamagnetic behavior at room 
temperature. The fitting of the magnetization curve to that of the 
calculation from the Langevin function revealed the distribution of particle 
sizes ranging from 2 nm to 5 nm. The TMR ratio exhibited an anomalous 
increase at low temperatures but no significant indication of change with 
bias voltage. In Ch. 4, we showed that the anomalous increase of the MR 
indicated evidence for higher-order tunneling (cotunneling) between large 
granules through intervening small granules. We emphasize that the existence 
of higher-order tunneling is a natural consequence of the granular 
structure, since broad distribution of granule sizes is an intrinsic 
property of granular systems.

In Ch. 5, we concentrated on SD-SET properties in a sequential tunneling 
regime. In order to study it, the tunnel paths between electrodes had to be 
restricted because the film contains a large number of particles of 
different sizes. We fabricated two types of device structures with Co-Al-O 
film using focused ion-beam milling or electron-beam lithography techniques. 
One is a granular nanobridge structure: point-shaped electrodes separated by 
a very narrow lateral gap filled with Co-Al-O granular film. The other is a 
current-perpendicular-to-plane (CPP) geometry structure: a thin Co-Al-O 
granular film sandwiched by ferromagnetic electrodes with the current 
flowing in the direction perpendicular to the film plane through a few Co 
particles. We measured the current-bias voltage curves in these samples, and 
found the enhancement and oscillation of TMR due to spin-dependent SET in a 
sequential tunneling regime. We made a theoretical calculation employing the 
orthodox theory of SET and explained the enhanced TMR just above the Coulomb 
threshold voltage. In Ch. 6, we reported experimental evidence of the spin 
accumulation effect in Co nanoparticles leading to the oscillation of TMR 
with alternate sign changes. Furthermore, the spin relaxation time in Co 
nanoparticles was also evaluated by comparing the experimental results to 
numerical simulations. We discovered that the spin relaxation time in the 
nanoparticles is unprecedentedly enhanced up to more than the order of 
hundreds of nanoseconds, compared to that evaluated from the spin diffusion 
length of ferromagnetic layers in previous CPP-GMR studies, $i.e., $the order of 
tens of picoseconds. 

Although the spin relaxation mechanism of the magnetic nanoparticle is not 
well understood at present, there are some candidates for cause of such a 
dramatic increase of the spin relaxation time: suppression of spin-flip 
scattering by spin-orbit interaction due to the discrete energy level 
spacing in the Co nanoparticle and/or blockade of spin-lattice relaxation 
due to the amorphous Al-O matrix. Further studies such as measuring the 
temperature dependence of spin relaxation time will clarify the mechanism of 
spin relaxation in nanoparticles.\newline

\textbf{Acknowledgements}

We thank H. Imamura (AIST, Tsukuba), S. Takahashi, J. Martinek, S. Maekawa 
(Tohoku Univ., Sendai), J. Inoue (Nagoya Univ.) for useful discussions. We 
also thank K. Yamane (Tohoku Univ., Sendai), N. Kobayashi, S. Ohnuma, T. 
Masumoto (RIEMM, Sendai), M. Ohnuma, K. Hono (NIMS, Tsukuba), S. Nagata 
(Tohoku Univ., Sendai) for helping with sample preparation, structural 
analysis and film composition analysis. Parts of the sample preparation were 
performed at the Advanced Research Center of Metallic Glasses, IMR, Tohoku 
University. This work was partially supported by CREST-JST.

\newpage 
\textbf{References}\newline

\textbf{[1] M. N. Baibich, J. M. Broto, A. Fert, F. N. Vandau, F. Petroff, 
P. Eitenne, G. Creuzet, A. Friederich, and J. Chazelas, Physical Review 
Letters 61 (1988) 2472.}

\textbf{[2] S. Maekawa and U. Gafvert, Ieee Transactions on Magnetics 18 
(1982) 707.}

\textbf{[3] T. Miyazaki and N. Tezuka, Journal of Magnetism and Magnetic 
Materials 139 (1995) L231.}

\textbf{[4] J. S. Moodera, L. R. Kinder, T. M. Wong, and R. Meservey, 
Physical Review Letters 74 (1995) 3273.}

\textbf{[5] E. Y. Tsymbal, O. N. Mryasov, and P. R. LeClair, Journal of 
Physics-Condensed Matter 15 (2003) R109.}

\textbf{[6] S. A. Wolf, D. D. Awschalom, R. A. Buhrman, J. M. Daughton, S. 
von Molnar, M. L. Roukes, A. Y. Chtchelkanova, and D. M. Treger, Science 294 
(2001) 1488.}

\textbf{[7] G. A. Prinz, Science 282 (1998) 1660.}

\textbf{[8] I. Zutic, J. Fabian, and S. Das Sarma, Reviews of Modern Physics 
76 (2004) 323.}

\textbf{[9] in Single Charge Tunneling, Vol. 294 (H. Gravert and M. H. 
Devoret, eds.), Plenum Press, New York, 1992.}

\textbf{[10] M. Julliere, Physics Letters A54 (1975) 225.}

\textbf{[11] J. Inoue and S. Maekawa, Physical Review B 53 (1996) 11927.}

\textbf{[12] D. AVERIN, MESOSCOPIC PHENOMENA (1991) 173.}

\textbf{[13] C. ADKINS, PHILOS MAG 37 (1977) 1285.}

\textbf{[14] J. Barnas and A. Fert, Physical Review Letters 80 (1998) 1058.}

\textbf{[15] J. Barnas and A. Fert, Europhysics Letters 44 (1998) 85.}

\textbf{[16] K. Majumdar and S. Hershfield, Physical Review B 57 (1998) 
11521.}

\textbf{[17] W. Wetzels, G. E. W. Bauer, and M. Grifoni, Physical Review B 
72 (2005) 020407.}

\textbf{[18] J. Martinek, J. Barnas, S. Maekawa, H. Schoeller, and G. Schon, 
Journal of Magnetism and Magnetic Materials 240 (2002) 143.}

\textbf{[19] J. Martinek, J. Barnas, S. Maekawa, H. Schoeller, and G. Schon, 
Physical Review B 66 (2002) 014402.}

\textbf{[20] J. Martinek, J. Barnas, A. Fert, S. Maekawa, and G. Schon, 
Journal of Applied Physics 93 (2003) 8265.}

\textbf{[21] A. Brataas, M. Hirano, J. Inoue, Y. V. Nazarov, and G. E. W. 
Bauer, Japanese Journal of Applied Physics Part 1-Regular Papers Short Notes 
{\&} Review Papers 40 (2001) 2329.}

\textbf{[22] A. Brataas and X. H. Wang, Physical Review B 6410 (2001) 
104434.}

\textbf{[23] A. Brataas, Y. V. Nazarov, J. Inoue, and G. E. W. Bauer, 
Physical Review B 59 (1999) 93.}

\textbf{[24] A. Brataas, Y. V. Nazarov, J. Inoue, and G. E. W. Bauer, 
Journal of Magnetism and Magnetic Materials 199 (1999) 176.}

\textbf{[25] A. Brataas, Y. V. Nazarov, J. Inoue, and G. E. W. Bauer, 
European Physical Journal B 9 (1999) 421.}

\textbf{[26] I. Weymann, J. Barnas, and J. Martinek, Journal of 
Superconductivity 16 (2003) 225.}

\textbf{[27] I. Weymann and J. Barnas, Physica Status Solidi B-Basic 
Research 236 (2003) 651.}

\textbf{[28] I. Weymann and J. Barnas, Physica Status Solidi B-Basic Solid 
State Physics 243 (2006) 239.}

\textbf{[29] J. Barnas, I. Weymann, J. Wisniewska, M. Kowalik, and H. W. 
Kunert, Materials Science and Engineering B-Solid State Materials for 
Advanced Technology 126 (2006) 275.}

\textbf{[30] I. Weymann and J. Barnas, Physical Review B 73 (2006) 205309.}

\textbf{[31] W. Rudzinski, J. Barnas, R. Swirkowicz, and M. Wilczynski, 
Physical Review B 71 (2005) 205307.}

\textbf{[32] W. Rudzinski and J. Barnas, Journal of Magnetism and Magnetic 
Materials 240 (2002) 124.}

\textbf{[33] W. Rudzinski and J. Barnas, Physical Review B 6408 (2001) 
085318.}

\textbf{[34] J. Martinek, J. Barnas, G. Michalek, B. R. Bulka, and A. Fert, 
Journal of Magnetism and Magnetic Materials 207 (1999) L1.}

\textbf{[35] M. Pirmann, J. von Delft, and G. Schon, Journal of Magnetism 
and Magnetic Materials 219 (2000) 104.}

\textbf{[36] in Spin Dependent Transport in Magnetic Nanostructures, Vol. 3 
(S. Maekawa and T. Shinjo, eds.), Taylor {\&} Francis, New York, 2002.}

\textbf{[37] I. Weymann and J. Barnas, Physical Review B 73 (2006) 033409.}

\textbf{[38] D. V. Averin and A. A. Odintsov, Physics Letters A 140 (1989) 
251.}

\textbf{[39] X. H. Wang and A. Brataas, Physical Review Letters 83 (1999) 
5138.}

\textbf{[40] S. Takahashi and S. Maekawa, Journal of the Physical Society of 
Japan 69 (2000) 102.}

\textbf{[41] S. Takahashi and S. Maekawa, Physical Review Letters 80 (1998) 
1758.}

\textbf{[42] K. Ono, H. Shimada, S. Kobayashi, and Y. Ootuka, Journal of the 
Physical Society of Japan 65 (1996) 3449.}

\textbf{[43] K. Ono, H. Shimada, and Y. Ootuka, Journal of the Physical 
Society of Japan 66 (1997) 1261.}

\textbf{[44] K. Ono, H. Shimada, and Y. Ootuka, Journal of the Physical 
Society of Japan 67 (1998) 2852.}

\textbf{[45] K. Ono, H. Shimada, and Y. Ootuka, Solid-State Electronics 42 
(1998) 1407.}

\textbf{[46] Y. Ootuka, R. Matsuda, K. Ono, and H. Shimada, Physica B 280 
(2000) 394.}

\textbf{[47] Y. Ootuka, K. Ono, H. Shimada, R. Matsuda, and A. Kanda, 
Materials Science and Engineering B-Solid State Materials for Advanced 
Technology 84 (2001) 114.}

\textbf{[48] K. Nakajima, Y. Saito, S. Nakamura, and K. Inomata, Ieee 
Transactions on Magnetics 36 (2000) 2806.}

\textbf{[49] F. Petroff, L. F. Schelp, S. F. Lee, F. Fettar, P. Holody, A. 
Vaures, J. L. Maurice, and A. Fert, Journal of Magnetism and Magnetic 
Materials 175 (1997) 33.}

\textbf{[50] T. Niizeki, H. Kubota, Y. Ando, and T. Miyazaki, Journal of 
Magnetism and Magnetic Materials 272-76 (2004) 1947.}

\textbf{[51] T. Niizeki, H. Kubota, Y. Ando, and T. Miyazaki, Journal of 
Applied Physics 97 (2005) 10C909.}

\textbf{[52] H. Sukegawa, A. Hirohata, S. Nakamura, N. Tezuka, S. Sugimoto, 
and K. Inomata, Ieee Transactions on Magnetics 41 (2005) 2679.}

\textbf{[53] H. Sukegawa, S. Nakamura, A. Hirohata, N. Tezuka, and K. 
Inomata, Physical Review Letters 94 (2005) 068304.}

\textbf{[54] F. Schelp, S. F. Lee, F. Fettar, F. N. vanDau, F. Petroff, A. 
Vaures, and A. Fert, Journal of Applied Physics 81 (1997) 5508.}

\textbf{[55] L. F. Schelp, A. Fert, F. Fettar, P. Holody, S. F. Lee, J. L. 
Maurice, F. Petroff, and A. Vaures, Physical Review B 56 (1997) R5747.}

\textbf{[56] Z. M. Zeng, X. F. Han, W. S. Zhan, Y. Wang, Z. Zhang, and S. 
Zhang, Physical Review B 72 (2005) 054419.}

\textbf{[57] F. Ernult, S. Mitani, K. Takanashi, Y. K. Takahashi, K. Hono, 
Y. Takahashi, and E. Matsubara, Applied Physics Letters 87 (2005) 033115.}

\textbf{[58] F. Ernult, K. Yamane, S. Mitani, K. Yakushiji, K. Takanashi, Y. 
K. Takahashi, and K. Hono, Applied Physics Letters 84 (2004) 3106.}

\textbf{[59] C. T. Black, C. B. Murray, R. L. Sandstrom, and S. H. Sun, 
Science 290 (2000) 1131.}

\textbf{[60] K. Yakushiji, S. Mitani, K. Takanashi, S. Takahashi, S. 
Maekawa, H. Imamura, and H. Fujimori, Applied Physics Letters 78 (2001) 
515.}

\textbf{[61] S. Mitani, H. Fujimori, K. Takanashi, K. Yakushiji, J. G. Ha, 
S. Takahashi, S. Maekawa, S. Ohnuma, N. Kobayashi, T. Masumoto, M. Ohnuma, 
and K. Hono, Journal of Magnetism and Magnetic Materials 199 (1999) 179.}

\textbf{[62] S. Mitani, S. Takahashi, K. Takanashi, K. Yakushiji, S. 
Maekawa, and H. Fujimori, Physical Review Letters 81 (1998) 2799.}

\textbf{[63] K. Yakushiji, F. Ernult, H. Imamura, K. Yamane, S. Mitani, K. 
Takanashi, S. Takahashi, S. Maekawa, and H. Fujimori, Nature Materials 4 
(2005) 57.}

\textbf{[64] K. Yakushiji, S. Mitani, K. Takanashi, and H. Fujimori, Journal 
of Physics D-Applied Physics 35 (2002) 2422.}

\textbf{[65] K. Yakushiji, S. Mitani, K. Takanashi, and H. Fujimori, Journal 
of Applied Physics 91 (2002) 7038.}

\textbf{[66] S. Mitani, K. Takanashi, K. Yakushiji, and H. Fujimori, Journal 
of Applied Physics 83 (1998) 6524.}

\textbf{[67] S. Mitani, K. Takanashi, K. Yakushiji, J. Chiba, and H. 
Fujimori, Materials Science and Engineering B-Solid State Materials for 
Advanced Technology 84 (2001) 120.}

\textbf{[68] A. Bernand-Mantel, P. Seneor, N. Lidgi, M. Munoz, V. Cros, S. 
Fusil, K. Bouzehouane, C. Deranlot, A. Vaures, F. Petroff, and A. Fert, 
Applied Physics Letters 89 (2006) 062502.}

\textbf{[69] R. Matsuda, A. Kanda, and Y. Ootuka, Physica B-Condensed Matter 
329 (2003) 1304.}

\textbf{[70] S. Haraichi and T. Wada, Journal of Applied Physics 95 (2004) 
7249.}

\textbf{[71] S. Haraichi, T. Wada, K. Ishii, and K. Hikosaka, Japanese 
Journal of Applied Physics Part 1-Regular Papers Short Notes {\&} Review 
Papers 43 (2004) 6061.}

\textbf{[72] Y. Nogi, H. Wang, F. Ernult, K. Yakushiji, S. Mitani, and K. 
Takanashi, Journal of Physics D-Applied Physics 40 (2007) 1242.}

\textbf{[73] J. Konig, H. Schoeller, and G. Schon, Physical Review Letters 
78 (1997) 4482.}

\textbf{[74] M. Johnson and R. H. Silsbee, Physical Review Letters 55 (1985) 
1790.}

\textbf{[75] F. J. Jedema, H. B. Heersche, A. T. Filip, J. J. A. Baselmans, 
and B. J. van Wees, Nature 416 (2002) 713.}

\textbf{[76] F. J. Jedema, A. T. Filip, and B. J. van Wees, Nature 410 
(2001) 345.}

\textbf{[77] J. Barnas and A. Fert, Journal of Magnetism and Magnetic 
Materials 192 (1999) L391.}

\textbf{[78] J. Inoue and A. Brataas, Physical Review B 70 (2004) 140406.}

\textbf{[79] J. GITTLEMAN, PHYSICAL REVIEW B 5 (1972) 3609.}

\textbf{[80] J. S. Helman and B. Abeles, Physical Review Letters 37 (1976) 
1429.}

\textbf{[81] H. Fujimori, S. Mitani, and S. Ohnuma, Materials Science and 
Engineering B-Solid State Materials for Advanced Technology 31 (1995) 219.}

\textbf{[82] S. Mitani, H. Fujimori, and S. Ohnuma, Journal of Magnetism and 
Magnetic Materials 177 (1998) 919.}

\textbf{[83] M. Ohnuma, K. Hono, E. Abe, and H. Onodera, Journal of Applied 
Physics 82 (1997) 5646.}

\textbf{[84] P. Sheng, B. Abeles, and Y. Arie, Physical Review Letters 31 
(1973) 44.}

\textbf{[85] K. Yakushiji, S. Mitani, K. Takanashi, J. G. Ha, and H. 
Fujimori, Journal of Magnetism and Magnetic Materials 212 (2000) 75.}

\textbf{[86] M. Ohnuma, K. Hono, H. Onodera, J. S. Pedersen, S. Mitani, and 
H. Fujimori, in Advances in Nanocrystallization, Vol. 307, 1999, p. 171.}

\textbf{[87] D. V. Averin, A. A. Odintsov, and S. V. Vyshenskii, Journal of 
Applied Physics 73 (1993) 1297.}

\textbf{[88] D. V. Averin, A. N. Korotkov, A. J. Manninen, and J. P. Pekola, 
Physical Review Letters 78 (1997) 4821.}

\textbf{[89] B. Abeles, Advances in Phyisics 24 (1975) 407.}

\textbf{[90] K. Takanashi, S. Mitani, J. Chiba, and H. Fujimori, Journal of 
Applied Physics 87 (2000) 6331.}

\textbf{[91] H. Imamura, J. Chiba, S. Mitani, K. Takanashi, S. Takahashi, S. 
Maekawa, and H. Fujimori, Physical Review B 61 (2000) 46.}

\textbf{[92] R. J. Soulen, J. M. Byers, M. S. Osofsky, B. Nadgorny, T. 
Ambrose, S. F. Cheng, P. R. Broussard, C. T. Tanaka, J. Nowak, J. S. 
Moodera, A. Barry, and J. M. D. Coey, Science 282 (1998) 85.}

\textbf{[93] M. Fujii, T. Kita, S. Hayashi, and K. Yamamoto, Journal of 
Physics-Condensed Matter 9 (1997) 8669.}

\textbf{[94] J. C. Slonczewski, Physical Review B 39 (1989) 6995.}

\textbf{[95] A. Brataas, Y. V. Nazarov, and G. E. W. Bauer, Physical Review 
Letters 84 (2000) 2481.}

\textbf{[96] D. Huertas-Hernando, Y. V. Nazarov, A. Brataas, and G. E. W. 
Bauer, Physical Review B 62 (2000) 5700.}

\textbf{[97] M. Beth Stearns, Journal of Magnetism and Magnetic Materials 5 
(1977) 167.}

\textbf{[98] R. Meservey and P. M. Tedrow, Physics Reports-Review Section of 
Physics Letters 238 (1994) 173.}

\textbf{[99] S. Gueron, M. M. Deshmukh, E. B. Myers, and D. C. Ralph, 
Physical Review Letters 83 (1999) 4148.}

\textbf{[100] M. M. Deshmukh, S. Kleff, S. Gueron, E. Bonet, A. N. 
Pasupathy, J. von Delft, and D. C. Ralph, Physical Review Letters 8722 
(2001) 226801.}

\textbf{[101] S. Kleff, J. von Delft, M. M. Deshmukh, and D. C. Ralph, 
Physical Review B 64 (2001) 220401.}

\textbf{[102] S. Kleff and J. von Delft, Physical Review B 65 (2002) 
214421.}

\textbf{[103] A. Kawabata, JOURNAL OF THE PHYSICAL SOCIETY OF JAPAN 29 
(1970) 902.}

\textbf{[104] G. G. Khaliullin and M. G. Khusainov, Zhurnal Eksperimentalnoi 
I Teoreticheskoi Fiziki 94 (1988) 163.}

\textbf{[105] G. Mitrikas, C. C. Trapalis, and G. Kordas, Journal of 
Chemical Physics 111 (1999) 8098.}

\textbf{[106] K. Yakushiji, S. Mitani, K. Takanashi, and H. Fujimori, J. 
Magn. Soc. Jpn 22 (1998) 577.}

\newpage 
\textbf{Figure captions}\newline

Fig. 2-1 Schematic diagram of a circuit including a single-electron 
tunneling (SET) double junction.\newline

Fig. 2-2 Current ($I)$ - bias voltage ($V)$ characteristics of a SET junction (a) 
with no asymmetry [of tunnel] resistances, and (b) with strong asymmetry of 
tunnel resistances.\newline

Fig. 2-3 Schematic diagram of a double tunnel junction to explain chemical 
potential shift due to spin accumulation in the ferromagnetic nanoparticle.\newline

Fig. 3-1 Schematic illustration of a sputtering method for preparing 
insulating granular films: (a) reactive sputtering, (b) sputtering with a 
composition target and (c) tandem deposition with plural targets.\newline

Fig. 3-2 (a) Plan view and (b) cross-sectional view of transmission electron 
microscopy (TEM) micrographs for a Co$_{46}$Al$_{19}$O$_{35}$ film.\newline

Fig. 3-3 High-resolution electron microscopy (HREM) micrograph for 
Co$_{36}$Al$_{22}$O$_{42}$ granular film.\newline

Fig. 3-4 Temperature ($T)$ dependence of electrical resistivity (\textit{$\rho $}) for Co-Al-O 
films of different compositions: Co$_{54}$Al$_{21}$O$_{25}$, 
Co$_{52}$Al$_{20}$O$_{28}$, Co$_{46}$Al$_{19}$O$_{35}$, and 
Co$_{36}$Al$_{22}$O$_{42}$.\newline

Fig. 3-5 (a) Magnetic-field dependence of magnetization ($M-H)$ and (b) 
temperature dependence of magnetization ($M-T)$ for the 
Co$_{36}$Al$_{22}$O$_{42}$ film. In (b), $M-T$ curve of field cool (FC) was 
measured at $H$ = 20 Oe.\newline

Fig. 3-6 Magnetic-field dependence of electrical resistivity (MR curve) 
measured at (a) room temperature and (b) 4.2 K. Magnetoresistance (MR) ratio 
shown in the left axis was estimated from the definition: MR = $\Delta 
\rho {\rm v}\rho _{max}$.\newline

Fig. 3-7 The result of particle size distribution of a 
Co$_{46}$Al$_{19}$O$_{35}$ film estimated from superparamagnetic behavior. 
(a) Result of fitting calculated magnetization curves to experimental 
results at $T$ = 200 and 293 K. (b) The size distribution which gives the best 
fit in (a).\newline

Fig. 4-1 Temperature ($T)$ dependence of the tunnel magnetoresistance (TMR) 
ratio for Co-Al-O films of various compositions: Co$_{54}$Al$_{21}$O$_{25}$, 
Co$_{52}$Al$_{20}$O$_{28}$, Co$_{46}$Al$_{19}$O$_{35}$, and 
Co$_{36}$Al$_{22}$O$_{42}$. Solid curves (a, b, c, and d) represent the 
theoretical magnetoresistance ratio given by Eq. (4.2) with spin 
polarization $P$; a: $P$ = 0.306; b: $P$ = 0.290; c: $P$ = 0.275; d: $P$ = 0.250. \newline

Fig. 4-2 Schematic illustrations of current-perpendicular-to-plane (CPP) 
geometry sample where a 1 $\mu $m thick Co$_{36}$Al$_{22}$O$_{42}$ granular 
film was sandwiched between upper and lower Au-Cr electrodes.\newline

Fig. 4-3 Bias voltage ($V_{b})$ dependence of $\rho $ at H = 0 (a) and 
$V_{b}$ dependence of TMR (b) for a CPP geometry sample with 
Co$_{36}$Al$_{22}$O$_{42}$ measured at 4.2 K. Closed circles represent the 
experimental results and solid curves represent the theoretical ones using 
Eq. (4.6) with$ T$ = 4.2 K and $N_{g}$ = 140.\newline

Fig. 4-4 (a) Schematic illustration of granular structure and a higher-order 
tunneling process where a charge carrier is transferred from the charged 
large granule (left), via the two small ones, to the neutral large one 
(right). (b) Model structure used for the calculation of conductivity.\newline

Fig. 4-5 Temperature ($T)$ dependence of electrical resistivity (\textit{$\rho $}) for Co-Al-O 
films of different compositions: Co$_{54}$Al$_{21}$O$_{25}$, 
Co$_{52}$Al$_{20}$O$_{28}$, Co$_{46}$Al$_{19}$O$_{35}$, and 
Co$_{36}$Al$_{22}$O$_{42}$. Solid lines represent the theoretical results 
given by Eq. (4.2) with charging energy $E_{c}$; Co$_{36}$Al$_{22}$O$_{42}$: 
$E_{c}$ / $k_{B}$ = 110 K; Co$_{46}$Al$_{19}$O$_{35}$:$ E_{c}$ / $k_{B}$ = 25 K; 
Co$_{52}$Al$_{20}$O$_{28}$: $E_{c}$ / $k_{B}$ = 18 K; 
Co$_{54}$Al$_{21}$O$_{25}$: $E_{c}$ / $k_{B}$ = 9 K.\newline

Fig. 5-1 (a) Scanning tunneling microscopy (STM) topographic image and (b) 
current -- bias voltage characteristics based on scanning tunneling 
spectroscopy (STS) measurements at room temperature for a 
Co$_{36}$Al$_{22}$O$_{42}$ film.\newline

Fig. 5-2 (a) Schematic view of an insulating granular nanobridge. (b) 
Scanning ion microscopy image of the NbZrSi electrodes separated by a 
nanometer-sized lateral gap. The length ($l)$ is 30 nm and the width ($w)$ is 60 
nm.\newline

Fig. 5-3 (a) Current -- bias voltage ($I-V_{b})$ characteristics and (b) 
$V_{b}$ dependence of TMR measured at 4.2 K for a sample of $w$ = 60 nm, $l$ = 30 
nm and $t$ = 7.5 nm. In (a), the solid and dashed curves represent 
$I-V_{b}$ curves in a magnified current range at $H$ = 0 and 10 kOe, 
respectively. $I-V_{b}$ curves throughout the measured current range are shown 
in the inset.\newline

Fig. 5-4 (a) Current -- bias voltage ($I-V_{b})$ characteristics and (b) 
$V_{b}$ dependence of TMR measured at 4.2 K for a sample of $w$ = 700 nm, $l$ = 40 
nm and $t$ = 15 nm. Viewgraphs are shown in the same form as in Fig. 5-3.\newline

Fig. 5-5 $V_{p}$ vs. $V_{th}$ for the various samples. $V_{p}$ is the voltage 
where the TMR shows the maximum.\newline

Fig. 5-6 (a) Schematic illustration and equivalent circuit of the 
calculation model. The conducting path in the contact is modeled by a 
parallel circuit of 20 triple-tunnel-junctions, assuming the tunnel 
resistance between an electrode and a particle ($R_{ep})$, tunnel resistances 
between particles ($R_{pp})$ and junction capacitances ($C_{ep}$ and 
$C_{pp})$. These parameters are assumed to be distributed around the mean 
values, $i.e.$, $R_{ep} =(1\pm \delta )\overline R _{ep} $, $R_{pp} =(1\pm \delta 
)\overline R _{pp} $, $C_{ep} =(1\pm \delta )\overline C _{ep} $, and 
$C_{pp} =(1\pm \delta )\overline C _{pp} $, where \textit{$\delta $} is the deviation from the 
typical value. (b) The numerical calculation results (solid curve) and 
experimental results (dashed curve). \newline

Fig. 5-7 Schematic illustrations of (a) the CPP geometry sample, and (b) the 
process of making a contact area by FIB etching.\newline

Fig. 5-8 (a) Current -- bias voltage ($I-V_{b})$ curves and (b) bias-voltage 
dependence of TMR measured at 4.2 K for a CPP geometry sample with a 0.5 x 
0.5 $\mu $m$^{2}$ contact area. In (a), the solid and dashed curves 
represent $I-V_{b}$ curves at $H$ = 0 and 10 kOe, respectively.\newline

Fig. 6-1 Schematic illustration of a sample with pillar structure, prepared 
for the current-perpendicular-to-plane (CPP) measurement. It consists of Al 
bottom electrode / Al-O / Co-Al-O granular film / Co top electrode / Pt 
microfabricated by electron-beam lithography and Ar ion milling process to 
reduce the contact area down to 0.4 x 0.4 $\mu $m. \newline

Fig. 6-2 Magnetotransport properties measured at 4.2 K in an Al / Al-O / 
Co-Al-O / Co sample of sub-micron sized area. (a) Current ($I)$ - bias voltage 
($V)$ curves and (b) $V$ dependence of the tunnel magnetoresistance (TMR). In (a), 
the gray and black solid curves represent a clear Coulomb staircase at zero 
magnetic field ($H$ = 0) and applied field ($H$ = 10 kOe), respectively. (c) The 
magnetic field dependence of electrical resistance (MR curves) at $V$ = 0.05 V 
and 0.12 V at which TMR shows negative and positive values, respectively.\newline

Fig. 6-3 Numerical calculation results considering the spin accumulation 
effect on TMR. The model assumes that a spin relaxation time in the 
ferromagnetic island is infinite and that a conductance bottleneck exists at 
the junction between the ferromagnetic island and the paramagnet electrode. 
(a) Schematic illustration of the model: paramagnet / ferromagnet / 
ferromagnet double tunnel junction. (b) $V$ dependence of the chemical 
potential splitting $\Delta E_F^\sigma $, which is caused by the spin 
accumulation. (c) Current ($I)$ - bias voltage ($V)$ curves in the anti-parallel 
(AP, dashed line) and parallel (P, solid line) alignments. (d) $V$ dependence 
of the tunnel magnetoresistance (TMR). \newline

Fig. 6-4 Numerical calculation results in the case of converse arrangement 
of the conductance bottleneck, i.e., $R_{2,\uparrow }^{P}{P} $ is fixed to 
be 25.0 G$\Omega $ and to be 10 times larger than $R_{1,\uparrow }^{P} 
$. (a) Schematic illustration of the model: paramagnet / ferromagnet / 
ferromagnet double tunnel junction. (b) $V$ dependence of the chemical 
potential spliting $\Delta E_F^\sigma $. (c) Current ($I)$ - bias voltage ($V)$ 
curves in the anti-parallel (AP, dashed line) and parallel (P, solid line) 
alignments. (d) $V$ dependence of TMR.\newline

Fig. 6-5 Surface plot of the TMR, which is calculated by the orthodox 
theory, as a function of $V$ and $\tau _{SF} $.\newline

Fig. 6-6 $V$ dependence of TMR for various values of the spin relaxation time 
$\tau _{SF} $: (a) 1 ns, (b) 10 ns, (c) 150 ns and (d) infinity.\newline

Fig. 6-7 $V$ dependence of TMR of [the] calculation result ($\tau _{SF} $ = 150 
ns) and experimental TMR. The TMR curve is well reproduced by choosing the 
spin relaxation time to be $\tau _{SF} $ = 150 ns.\newline

\end{document}